\documentclass[article]{aa} 
\usepackage[varg]{txfonts}
\usepackage{graphicx}
\usepackage{ulem}
\usepackage{txfonts}
\usepackage{natbib}
\usepackage{multirow}
\usepackage[usenames]{color}

\usepackage{enumerate}
\usepackage{booktabs}

\usepackage{amsmath}
\usepackage[version=3]{mhchem} 

\usepackage[english]{babel}

\newcommand{\kevap}{k_{\rm evap}}
\newcommand{\kIR}{k_{\rm IR}}
\newcommand{\Eintra}{E^{\rm intra}}
\newcommand{\Einter}{E^{\rm inter}}
\newcommand{\Etot}{E^{\rm tot}}
\newcommand{\Omintra}{\Omega^{\rm intra}}
\newcommand{\Ominter}{\Omega^{\rm inter}}
\newcommand{\Etr}{\epsilon_{\rm{tr}}}
\newcommand{\Etrmin}{\epsilon_{\rm{tr}}^{\rm{min}}}

\newcommand{\ch}[2]{$\rm{C}_{#1}\rm{H}_{#2}$}
\newcommand{\agr}[4]{$(\rm{C}_{#1}\rm{H}_{#2})_{#3}^{#4}$}
\usepackage{color}
\begin{document}

\title{Absolute evaporation rates of non-rotating neutral PAH clusters   
      \thanks{Appendices are only available in electronic form at http://www.aanda.org}
      } 

   \author{J. Montillaud\inst{1,2,}\thanks{\textit{Present address:}
         Institut Utinam, CNRS UMR 6213, OSU THETA, Université de Franche-Comté, 41bis avenue de l'Observatoire, 25000 Besançon, France} 
         \and
         C. Joblin\inst{1,2}
         }
 
   \institute{
   Universit\'e de Toulouse, UPS-OMP, IRAP, Toulouse, France
   \and
   CNRS, IRAP, 9 Av. colonel Roche, BP 44346, 31028, Toulouse Cedex 4, France
   }

\date{Received 28 November 2013; Accepted 17 Mars 2014}
\offprints{J.~Montillaud, \email{julien@obs-besancon.fr}}


   \abstract
   { Clusters of polycyclic aromatic hydrocarbons (PAHs) have been proposed as candidates for evaporating very small 
   grains, which are thought to be precursors of free-flying PAHs. Evaporation rates have been calculated so far 
   only for species containing up to a few 100-C atoms, whereas interstellar PAH clusters could contain up to $\sim1000$ 
   C atoms.}
   { We present a method that generalises the calculation of the statistical evaporation rate of large PAH clusters and
   provides rates for species containing up to $\sim1000$ C-atoms.}
   { The evaporation of non-rotating neutral homo-molecular PAH clusters containing up to 12 molecules from a 
   family of highly symmetric compact PAHs is studied. Statistical calculations were performed and completed with molecular 
   dynamics simulations at high internal energies to provide absolute values for the evaporation rate and distributions 
   of kinetic energy released. The calculations used explicit atom-atom Lennard-Jones potentials in the rigid molecule 
   approximation. A new method is proposed to take both inter- and intra-molecular vibrations into account.}
   {Without any parameter adjustment, the calculated evaporation rates agree well with available experimental data. We 
   find that the non-rotation assumption has a limited impact on the evaporation rates. The photostability of PAH 
   clusters increases dramatically with the size of molecules in the clusters, and to a lesser extent with the number of 
   molecules in the clusters. For values of the UV radiation field that are typical of the regions where evaporating 
   very small grains are observed, the smallest clusters in this study ($\sim 50$ C-atoms) are found to be quickly 
   photo-evaporated, whereas the largest clusters ($\sim 1000$ C-atoms) are photostable.
   }
   {Our results support the idea that large PAH clusters are good candidates for evaporating very small grains.}


   \maketitle
%


\section{Introduction}\label{sec:introduction}

   Astronomical observations have revealed a set of emission bands in the mid-infrared range, the so-called infrared 
   aromatic bands (AIBs), the most intense falling at 3.3, 6.2, 7.7, 8.6, 11.3, and 12.7 $\mu$m. Following the studies of 
   \citet{leger_identification_1984} and \citet{allamandola_polycyclic_1985}, this emission is generally attributed to 
   polycyclic aromatic hydrocarbons (PAHs), and this proposal has motivated numerous studies on PAHs \citep[see][for a 
   recent compilation of works on this topic]{joblin_pahs_2011}. 
   
   \citet{rapacioli_spectroscopy_2005} have analysed the spectral variations of the AIBs in photodissociation regions (PDRs)
   and suggest that free PAHs are produced by evaporation of very small grains (VSGs) containing at least ${\sim400}$ 
   carbon atoms. They propose that VSGs are made of PAH clusters that evaporate under UV irradiation. \citep
   {pilleri_evaporating_2012} confirm this mechanism over a wide range of physical conditions and propose to call 
   these species evaporating very small grains (eVSG).
   
   PAH clusters are also considered as key species in flames and circumstellar shells, where they could be involved 
   in the transition from molecular growth to nucleation of soot particles \citep{frenklach_reaction_2002, 
   cherchneff_formation_2011}. The dimerisation of pyrene (\ch{16}{10}) and sometimes of other PAHs has been invoked in 
   numerous studies \citep[][and references therein]{schuetz_nucleation_2002} as a critical step in soot formation 
   models. However, the experimental results of \citet{sabbah_exploring_2010} contradict this proposal unless larger 
   species than pyrene are involved.

   The size distribution of PAH clusters in natural environments, such as flames, PDRs, or circumstellar shells, results 
   from the balance between nucleation and evaporation processes. To our knowledge there is only one experimental study 
   of the evaporation properties of large PAH clusters by \citet{schmidt_coronene_2006}, while several theoretical 
   studies address the questions of their structure and stability \citep{rapacioli_stacked_2005,herdman_intermolecular_2008}. 
   In addition, \citet{rapacioli_formation_2006} initiated a theoretical investigation on the formation and photo-evaporation 
   in PDRs of neutral clusters containing up to $\sim 300$ C-atoms. They show that in regions where eVSGs are observed, 
   these clusters are photo-evaporated much faster than they can be reformed by collisions. However, this study was based on 
   coronene (\ch{24}{12}), whereas we have recently shown that PAHs smaller than $\sim50-70$ carbon atoms are strongly 
   photo-dissociated, and therefore larger PAHs have to be considered \citep{montillaud_evolution_2013}.
   
   In this study, we aim at providing absolute evaporation rates for neutral PAH clusters of large molecules. These
   rates are calculated as a function of the total excess energy in the clusters, defined as the sum of inter- and 
   intra-molecular potential energy and of kinetic energy. To sample the parameter space of PAH cluster size, we 
   focussed on three highly symmetric and compact species, namely coronene \ch{24}{12}, circumcoronene \ch{54}{18}, and 
   circumcircumcoronene \ch{96}{24}, forming clusters containing up to 12 molecules. For the sake of simplicity, 
   we limit the study to non-rotating clusters. We show in Sect.~\ref{sec:PSTinput} and Appendix~\ref{anx:sensitivity} 
   that this assumption has no strong impact on the applicability of our results in the astrophysical context.
   
   The calculations of absolute evaporation rates by \citet{rapacioli_formation_2006} were based on the 
   statistico-dynamical method \citep{weerasinghe_absolute_1993, calvo_statistical_2003}. It consists in combining the 
   assets of (i) the phase space theory (PST), which provides accurate relative data over broad energy ranges and (ii)
   molecular dynamics (MD) simulations to get absolute data, but only for a limited range at high energies. When applied 
   to the evaporation of large molecular clusters, the need for long MD trajectories usually makes it necessary to simplify 
   the simulations by assuming that molecules are rigid. This approximation raises difficulties for taking 
   both the intra- and inter-molecular vibrations into account when combining PST and MD results. We propose here a simple 
   method for circumventing this difficulty while relaxing the rigid molecule approximation to some extent. To assess the 
   reliability of this method and of obtained molecular data, we compare our results on coronene clusters with the 
   experimental data by \citet{schmidt_coronene_2006}.
   
   The paper is structured as follows. The methods used to compute the structures, energetic properties, and evaporation 
   rates of PAH clusters are presented in the next section. Further details on the methods can be found in the appendix. 
   In Sect.~\ref{sec:result_disc}, we report our absolute evaporation rates and compare them with the available 
   experimental results. The first astrophysical consequences of our new molecular data are discussed in Sect.~\ref
   {sec:appli_astro}. A summing-up of this work with perspectives on possible extensions of our method, the remaining open 
   questions, and astrophysical modelling concludes the paper. Readers mostly interested in using the molecular data 
   and in their astrophysical consequences may skip the following section dedicated to methods.

\section{Methods}\label{sec:method}

\subsection{Structures and binding energies of neutral PAH clusters} 
\label{sec:structure}

   Prior to investigating the dynamical behaviour of the evaporation process, the knowledge of the static properties
   of the clusters, namely their ground state structure and binding energy, is required. \citet{rapacioli_stacked_2005}
   determined these properties for several neutral PAH clusters ranging from pyrene to circumcoronene clusters with 2
   to 32 molecules. They assumed the molecules to be rigid and to interact through an explicit atom-atom potential
   built on two contributions:

   \begin{equation}
      \label{eq:Vinter}
      V = \sum_{i<j} \sum_{\alpha \in i} \sum_{\beta \in j} \left[ V_{\rm LJ}(r_{i_{\alpha},j_{\beta}}) + V_{\rm Q}(r_{i_{\alpha},j_{\beta}}) \right]
   ,\end{equation}

   \noindent
   where $V_{\rm LJ}(r_{i_{\alpha},j_{\beta}})$ denotes the dispersion-repulsion energy between the atom $i_{\alpha}$
   of the molecule $i$ and the atom $j_{\beta}$ of the molecule $j$, and for which the Lennard-Jones (LJ) form is used. Here, 
   $V_{\rm Q}(r_{i_{\alpha},j_{\beta}})$ denotes the point charge electrostatic interaction between the partial charges 
   carried by the atoms of the molecules.
   
   For $V_{\rm LJ}$, we used the parameters proposed by \citet{van_de_waal_calculated_1983}.
   The electrostatic-potential-fitted (EPF) partial charges are used, because they exhibit good transferability from
   one PAH to another, according to \citet{rapacioli_stacked_2005}. In this work, for the coronene and circumcoronene 
   molecules, we used the structure and EPF charges computed by \citet{rapacioli_stacked_2005}. The structure and EPF 
   charges of the circumcircumcoronene molecule were derived from the B3LYP density functional theory (DFT) calculation
   of \citet{bauschlicher_infrared_2008}.

   \citet{rapacioli_stacked_2005} performed the optimisation with two methods: the basin-hopping or Monte-Carlo
   + minimisation method for the smaller clusters, and parallel tempering Monte-Carlo for the larger ones. They show 
   that for pyrene, coronene, and circumcoronene clusters, the lowest energies are obtained for single stacks in the 
   case of small numbers of molecules and for several stacks above 7, 8, and 16 molecules for pyrene, coronene, and 
   circumcoronene clusters, respectively. This strongly 
   suggests that the one-stack geometry prevails for circumcircumcoronene clusters containing up to at least 16 
   molecules. Therefore, we used the structures of coronene clusters computed by \citet{rapacioli_stacked_2005} as 
   first guess structures for \agr{96}{24}{N}{} (N=2,12). We then optimised these structures using classical molecular 
   dynamics simulations in the rigid molecule approximation using the following method. We computed several trajectories 
   with several initial kinetic energies between 0.01 and 2 eV in order to let the system explore the configurational 
   space around the initial structure, and we recorded the potential energy minima on-the-fly. Quenching was performed 
   for each recorded minimum, and we considered the structure corresponding to the deepest potential energy as optimised.
   
   Starting with a modified one-stack structure, we checked on the example of coronene clusters that our procedure leads 
   to the same geometries and energies as those determined by \citet{rapacioli_stacked_2005}. We found 
   circumcircumcoronene clusters to have the same one-twisted-stack geometries as coronene clusters. The 
   distance between two successive molecules is found to decrease from coronene (3.52 \AA) to circumcoronene (3.49 \AA) 
   and to circumcircumcoronene clusters (3.47 \AA). The same $\pi/6$ angle is found between two successive molecules. 
   \citet{herdman_intermolecular_2008} also studied the structures of neutral PAH dimers with a very similar approach 
   differing essentially through the choice of the point charge description. Their results favoured parallel-displaced 
   geometries as in graphite. However, \citet{rapacioli_stacked_2005} show that the final structure is very 
   sensitive to the point charge values, while the difference in energy between twisted and parallel-displaced stacks is 
   lower than 0.02 eV for the dimer of coronene. 
   
   \begin{figure}
      \begin{center}
         \includegraphics[angle=270, width=0.49\textwidth, keepaspectratio]{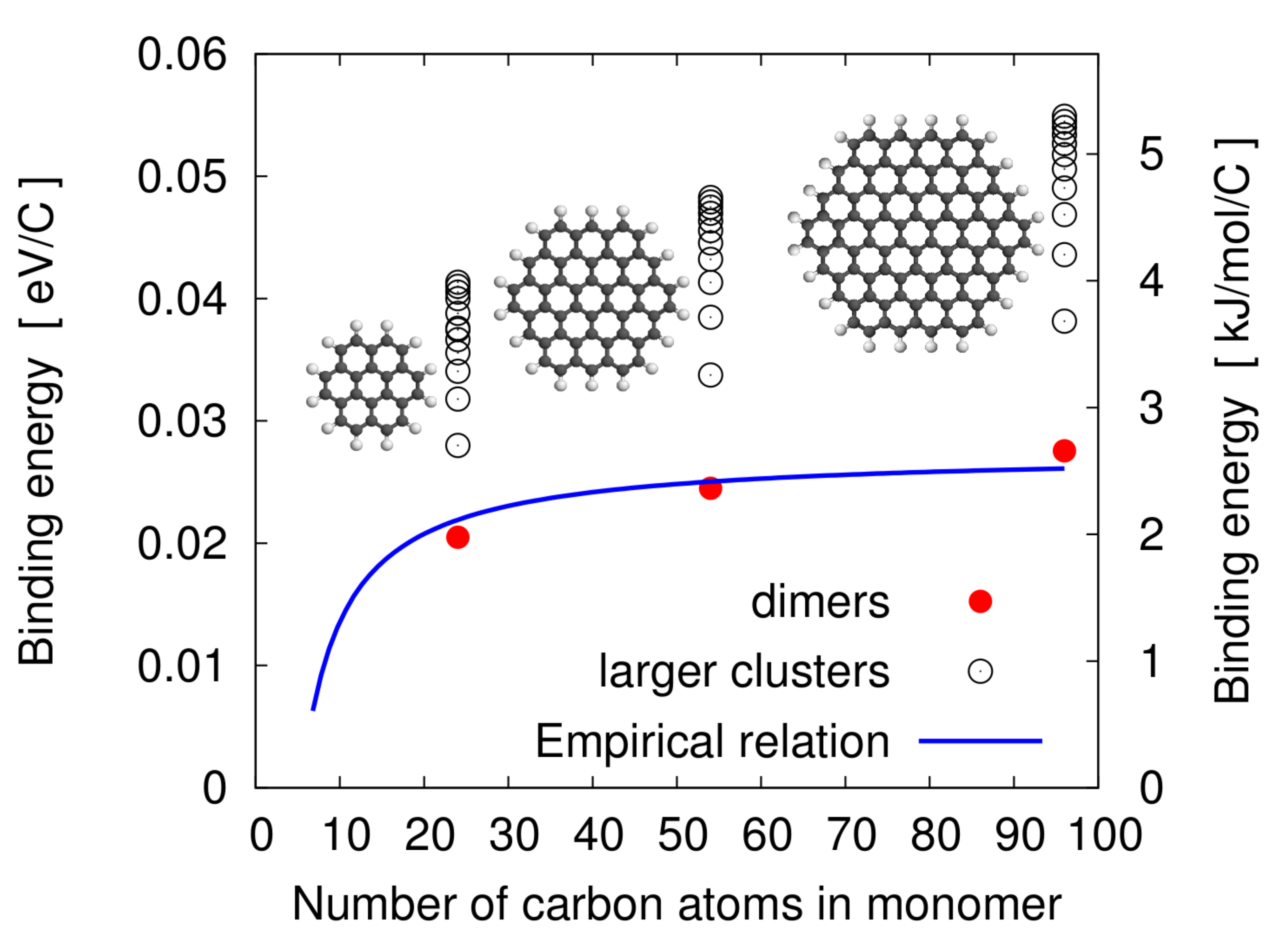}
      \end{center}
      \caption{ Binding energy per carbon atom in the cluster for some neutral homo-molecular PAH clusters. Red 
      filled circles are for dimers, black open circles for larger clusters with 3 to 12 molecules. The binding energy 
      per carbon atom increases with the size of monomers and with the number of molecules in the cluster. The blue 
      curve is the empirical relation determined by \citet{herdman_intermolecular_2008} and shows a maximum relative 
      difference of 7\% with our results. The structures of the three monomers considered in this study, coronene 
      \ch{24}{12}, circumcoronene \ch{54}{18}, and circumcircumcoronene \ch{96}{24}, are also shown.}
      \label{fig:Eliaison}
   \end{figure}

   In terms of binding energy, our results agree within 7\% with those of \citet{herdman_intermolecular_2008} 
   for the dimers. The binding energy increases faster than the number of carbon atoms in the monomers (see 
   Fig.~\ref{fig:Eliaison}) because the repulsive part of the intermolecular potential is related to the Coulomb interaction
   of peripheral atoms, whose number decreases relative to the total number of carbon atoms when the size of the 
   molecule increases. Because of long-range interactions, the binding energy per molecule increases with the number
   of molecules in the cluster. However, this effect is close to saturation for twelve molecules, giving an estimate of
   (i) the spatial extent of the intermolecular interactions and (ii) a typical size for bulk properties to emerge.

\subsection{Evaporative trajectories with molecular dynamics}\label{sec:MD}

   A first evaluation of the absolute evaporation rates was performed using molecular dynamics (MD) simulations for 
   non-rotating parent clusters. We considered molecules as rigid, and assumed they are frozen in their optimal structure. 
   This approximation prevents us from taking the coupling between intra- and inter-molecular vibrations into account. 
   However, only the 
   lowest frequency modes are expected to contribute significantly to this coupling, and full-dimensional calculations 
   are beyond reach for the large systems we are considering here. \citet{rapacioli_vibrations_2007} investigated the 
   effect of the rigid molecule approximation on the frequencies of intermolecular modes in coronene clusters. They 
   show that these frequencies are affected by less than 5\% for one-stack clusters, but larger differences are found 
   for two-stack structures. Therefore the results of our MD simulations should not be significantly affected by the 
   coupling between inter- and intra-molecular vibrations except for \agr{24}{12}{12}{}, which is the only two-stack 
   cluster in this work.
   
   The classical equations of motion were integrated using a leapfrog algorithm with quaternion coordinates, the forces
   between molecules being derived from the intermolecular potential presented in \ref{sec:structure}. The clusters were
   initially heated to a temperature $\sim$ 10 K along 1000 integration steps to prepare random initial conditions and 
   then suddenly heated to a given intermolecular energy. This was done by randomly distributing the excess energy in 
   molecular velocities during a single time step. The total linear and angular momenta were set to zero since our study 
   focusses on non-rotating clusters. The trajectories were propagated with time steps between 1 and 5 fs according to 
   the size of the molecules to ensure an energy conservation better than 0.1\% along the 10$^6$ steps ($\sim10^{-9} 
   s$) of each trajectory. The simulation was stopped after 10$^6$ integration steps or when an evaporation event was 
   observed, i.e. when the distance between one molecule and the mass centre exceeded five times the typical dimension of 
   the cluster. The last step for which the velocity of the evaporating molecule pointed inwards was considered as the 
   evaporation time. For each cluster, we computed enough trajectories to record 10000 evaporation events. For high 
   energies, each trajectory led to an evaporation, while for the lowest energies computed here, typically one 
   evaporation was recorded every two or three trajectories. We never observed the simultaneous evaporation of several 
   molecules. This is consistent with the dissociation energy being higher for such events than for single molecule 
   evaporations. The number $N(t)$ of trajectories that are not yet evaporated after time $t$ is expected to follow a 
   first-order kinetics law $N(t)=N(0)\exp(-k t)$, from which we extracted the evaporation rate $k$ using a linear fit 
   of the logarithmic variations in $N(t)$. 
   
   Evaporation rates were found to increase drastically with the excess energy in the cluster, defined here as the 
   sum of intermolecular potential energy and kinetic energy. The usual computational resources enable only high energy 
   simulations for which the evaporation rates are greater than $\sim10^{9} s^{-1}$. A statistical approach is needed to
   calculate evaporation rates at lower energies.

\subsection{Statistical approach of evaporation}\label{sec:thD}

 Phase space theory (PST) is a statistical theory that is well suited to describing molecular and cluster evaporation. It is 
presented in detail in several papers \citep[see for instance][]{chesnavich_statistical_1977, weerasinghe_absolute_1993, 
calvo_statistical_2003}. In this section, we are limited to the formalism needed to describe the evaporation of PAH 
clusters. We emphasise the difficulties raised by the use of the rigid molecule approximation and propose a simple 
method that enables these difficulties to be circumvented.

\subsubsection{The phase space theory framework}

   PST was shown to provide an efficient framework for computing unimolecular
   reaction rates \citep{weerasinghe_absolute_1993,calvo_statistical_2004,calvo_accurate_2010}. As 
   a statistical theory, it relates the evaporation rate to the ratio of the densities of states (DOS) of the fragments
   over the parent, with the specificity to take the conservation of both energy and angular momentum into account. 
   During the evaporation process, the initial angular momentum $J$ is shared between the rotation of the fragments 
   $J_r$ and the orbital angular momentum $L$. Similarly, the initial excess energy $E_N$ of the parent is shared 
   between the kinetic energy released (KER) $\Etr$, the internal energy of the released molecule $E_1$, the internal 
   energy of the fragment cluster $E_{N-1}$, and the dissociation energy $D_0^J$ that depends on $J$. Therefore, the 
   total DOS of the fragments results from the convolution of the vibrational densities of states (VDOS) 
   of the two fragments $\Omega_{N-1}$ and $\Omega_1$ and of the rotational density of states (RDOS) $\Gamma(\Etr,J)$.
   
   In the orbiting transition state version of PST \citep{chesnavich_statistical_1977}, the fragments separate with a 
   certain amount of translational-rotational energy $\Etr$ whose translational part $\epsilon_r$ has to overcome the 
   centrifugal barrier $\epsilon^{\dagger}(L)$ that arises from the centrifugal potential combined to the presence of a 
   long range interaction potential between the fragments. Therefore a minimum translational-rotational energy $\Etrmin$
   is required to enable the evaporation. The RDOS of the fragments $\Gamma(\Etr,J)$ results from the integration over 
   the accessible volume of rotational states whose boundaries emerge from the interplay between the conservation of 
   energy and angular momentum and depend on the geometry of the fragments. Detailed procedures for this calculation 
   can be found in the work by \citet{chesnavich_statistical_1977} and \citet{calvo_statistical_2004}. Finally, most
   of the complexity related to the conservation of angular momentum is encapsulated in the RDOS, and the evaporation 
   rate $\kevap$ in the PST framework can be  summarised by the relation:

   \begin{multline}
      \label{eq:PST-general}
         \kevap(E_N,J) = C_0 \, \int_{\Etrmin}^{E_N-D_0^J} d\Etr \int_{0}^{E_N-D_0^J-\Etr} dE_1 \\
         \times \, 
         \frac
         {
         \Omega_{N-1}(E_N-D_0^J-\Etr-E_1) ~~ \Omega_{1}(E_1) ~~ \Gamma(\Etr,J)
         }{
         \Omega_N(E_N-E_r)
         }
   \end{multline}

   \noindent  where $C_0$ is a constant, $E_r$  the rotational energy of the parent, $E_1$  the energy of the released
   molecule, $E_N-D_0^J-\Etr-E_1 = E_{N-1}$ is the energy of the larger fragment, and $E_N=\Einter_N + \Eintra_N$  the
   energy of the parent cluster, which is conserved during the evaporation process.
   
   Evaluating $C_0$ from theoretical considerations is not straightforward. For this reason it is customary to use
   MD simulations at high excess energy to calibrate $C_0$. We used the results of the MD simulations of 
   Sect.~\ref{sec:MD} for this purpose. However, MD calculations were performed using the rigid molecule approximation
   and therefore provide rigid molecule evaporation rates $\kevap^{\text{rigid}}$ as a function of 
   the intermolecular energy $\Einter$ of the parent clusters. In contrast, Eq.~\ref{eq:PST-general} provides evaporation 
   rates $\kevap$ considering all vibrational degrees of freedom, as a function of the total excess energy $E_N$ of the 
   parent clusters. It is therefore necessary to find a relationship between $\kevap$ and $\kevap^{\text{rigid}}$ before an
   estimate of $C_0$ can be given.

\subsubsection{To the rigid molecule approximation and beyond}\label{sec:rigid}

   In previous work, \citet{rapacioli_formation_2006} linked $\kevap^{\text{rigid}}$, expressed as a function of the 
   intermolecular energy $\Einter$, and $\kevap$, expressed as a function of the total internal energy in the cluster, 
   by computing the mean intermolecular energy for each total energy and exchanging the variables. This approach 
   neglects that the values of the intermolecular energy follow a broad distribution. We denote ${\cal P}_N^{\rm inter}
   (\Einter_N,E_N)$ as the density of probability for a cluster of $N$ molecules bearing an total energy of $E_N$ to have 
   an intermolecular energy $\Einter_N$. This quantity can be computed from the VDOS:

   \begin{equation}
      \label{eq:distrib-Einter}
      {\cal P}_N^{\rm inter}(\Einter_N,E_N) = \frac{\Ominter_N(\Einter_N) ~~ \Omintra_N(E_N - \Einter_N)}{\Omega_N(E_N)}
   \end{equation}

   \noindent where $\Ominter_N$ and $\Omintra_N$ are the VDOS of inter- and intra-molecular modes. Despite 
   their rare occurrence, the highest reachable values of $\Einter_N$ can play a major role in the evaporation process, 
   in particular for moderate values of the total energy for which the corresponding mean intermolecular energy is less 
   than the dissociation energy. Such situations can severely affect the lifetime of these species in interstellar 
   environments.
   
   We show here how to adapt Eq.~(\ref{eq:PST-general}) to the rigid molecule approximation, taking the 
   distribution of intermolecular energy into account. One needs to distinguish between inter- and intra-molecular modes. This can 
   be done by noticing that for a cluster with total energy $E$, the VDOS can be expanded as
   \begin{equation}
      \Omega(E) = \int_0^{E} \Ominter(\Einter) \Omintra(E-\Einter) d\Einter
   .\end{equation}
   We apply this procedure simultaneously to $\Omega_N$ and $\Omega_{N-1}$ in Eq.~\ref{eq:PST-general}, in order to explicitly 
   develop the integration over $\Einter_N$:
      \begin{multline}
      \label{eq:PST-integ-einter}
      \kevap(E_N,J) = 
      C_0  \times 
      \int_{\Etrmin}^{E_N-D_0^J} d\Etr 
      \int_{0}^{E_N-D_0^J-\Etr} dE_1
      \int_0^{E_N-E_r} d\Einter_N
      \\
      \times \, {\cal P}_N^{\rm inter} (\Einter_N,E_N-E_r) \\
      \times \, 
      \frac
      {
      \Gamma(\Etr,J) ~~ \Ominter_{N-1}(\Einter_{N-1}) ~~ \Omintra_{N-1}(\Eintra_{N-1}) ~~ \Omega_1(E_1)
      }
      {
      \Ominter_N(\Einter_N) ~~ \Omintra_N(\Eintra_N)
      }
      \end{multline}
   
   \noindent where the conservation of energy $E_r+\Einter_N+\Eintra_N=\Einter_{N-1}+E_1+\Eintra_{N-1}+D_0+\Etr$ {\it a 
   priori} prevents us from separating the integrations over intra- and inter-molecular energies. Nevertheless, in the 
   framework of the rigid molecule approximation, only intermolecular modes contribute to the dissociation process, and 
   the conservation of energy applies to inter- and intra-molecular energies independently, leading to $E_r+\Einter_N=
   \Einter_{N-1}+D_0+\Etr$. Therefore, the integrations over the intra- and inter-molecular energies can now be 
   performed separately. The integration over the intramolecular energy vanishes, while the rigid molecule evaporation 
   rate emerges from the integration over the KER distributions, leading to
   
   \begin{multline}
      \label{eq:kevap-link}
      \kevap(E_N,J) = \\
      \int_0^{E_N-E_r} {\cal P}_N^{\rm inter}(\Einter_N,E_N-E_r) \,
      \kevap^{\rm rigid}(\Einter_N+E_r,J) ~d\Einter_N
   \end{multline}

   \begin{multline}
      \label{eq:kevap-rigid}
      \kevap^{\rm rigid}(E=\Einter_N+E_r,J) = \\
      C_0 \int_{\Etrmin}^{E-D_0^J} 
      \frac
      {
      \Ominter_{N-1}(E-D_0^J-\Etr) ~~ \Gamma(\Etr,J)
      }{
      \Ominter_N(E-E_r)
      } d\Etr.
   \end{multline}

   The evaporation rate $\kevap(E_N,J)$ is therefore a simple statistical weighting of the rigid
   molecule evaporation rate, and the calibration constant $C_0$ is not modified through this procedure. Our method enables taking advantage of the rigid molecule approximation and recovering the major effects 
   of intramolecular vibrations afterwards in terms of statistical redistribution of the energy.

\subsection{Vibrational densities of states}\label{sec:VDOS}

   The accuracy of the calculated (microcanonical) evaporation rates depends on the accuracy of the vibrational
   densities of states. Computing anharmonic quantum VDOS for such large systems is beyond the scope of this work. We
   computed all other combinations of quantum + harmonic, classical + harmonic, and classical + anharmonic, and we
   specifically selected the most relevant for each term in Eq.~(\ref{eq:kevap-link}). The details of the
   calculations are provided in the appendix, along with the motivations for the chosen combinations. We summarise here
   which approximations were used for the VDOS involved in the calculations of the rates and distribution
   probabilities in the following.
   
   In Eq.~\ref{eq:kevap-rigid}, $\Ominter_{N}$ and $\Ominter_{N-1}$ were computed using the classical anharmonic 
   approximation. This choice provides the best possible description of the anharmonicity, which is important 
   at the high energies of the MD simulations (Sect.~\ref{sec:MD}).
   
   In Eq.~\ref{eq:distrib-Einter}, the whole range of energy is important. Because of the large number of 
   vibrational degrees of freedom in PAH clusters, the average energy per vibrational mode is expected to be low in most
   cases, therefore quantum effects are important. Thus, in this equation, all VDOS were computed using the 
   quantum harmonic approximation.

\section{Results and discussion}
\label{sec:result_disc}

   In this paper, we focus on the absolute evaporation rates of a few non-rotating PAH clusters, namely \agr{24}{12}{N}{}, 
   \agr{54}{18}{N}{} ($N=2,3,4,8,12$), and \agr{96}{24}{N}{} ($N=4,8$). To test our method, the results for coronene 
   clusters are compared with the experimental results of \citet{schmidt_coronene_2006}.

\subsection{Evaporation rates at fixed energy}
\label{sec:kevap}

   In this section, we discuss the parameters used to compute RDOS. The RDOS 
   are then combined with (i) the classical anharmonic intermolecular VDOS presented in Sect.~\ref{sec:VDOS_cl_anh} to 
   provide the rigid molecule evaporation rates, according to Eq.~(\ref{eq:kevap-rigid}), (ii) the MD results 
   obtained in Sect.~\ref{sec:MD} to derive absolute values of $\kevap^{\text{rigid}}$, and (iii) the probability 
   distribution of the intermolecular energy, according to Eq.~(\ref{eq:kevap-link}), to provide absolute evaporation 
   rates as a function of the total internal energy. The reliability of the method is assessed from 
   the comparison with MD data in Sect.~\ref{sec:PST_assessment}, and the absolute evaporation rates are presented in 
   Sect.~\ref{sec:PST_results}.

\subsubsection{PST input parameters}\label{sec:PSTinput}

   The typical evolution of the RDOS at low angular momenta is approximated well \citep{klots_reformulation_1971,
   calvo_statistical_2004} by $\Gamma(\Etr,J\approx0)\propto \Etr^{(r-1)/2}$, with $r=6$ being the number of rotational 
   degrees of freedom of the fragments. For the large species considered in this study, these variations are rather slow 
   with respect to the VDOS that scale with $E^{s-1}$ where $s$ is the number of vibrational degrees of freedom. 
   This leads to a low sensitivity of the evaporation rates to rotational parameters. Therefore, limiting this study to 
   non-rotating clusters does not exclude applying its results to clusters with rotational temperatures up to a 
   few 100 K, as in interstellar conditions \citep{ysard_long-wavelength_2010}.

   The effects of the geometry of the fragments on the KER distribution were discussed in detail 
   by Calvo and Parneix in a series of papers \citep{calvo_statistical_2003,parneix_statistical_2003,calvo_statistical_2004,
   parneix_statistical_2004}. \citet{rapacioli_formation_2006} considered the accuracy of the description of the cluster 
   geometry as a limitation to the study of neutral coronene clusters and limited their investigation to two clusters, 
   \agr{24}{12}{4}{} and \agr{24}{12}{13}{}, whose fragments are modelled well by a spherical cluster + oblate molecule 
   geometry. However, we found no quantitative argument for such a limitation and noticed that large errors on the 
   rotational constants of the fragments lead to very small variations in the KER distribution and the evaporation rate
   (see Appendix~\ref{anx:sensitivity}). Therefore, in the following, the parent and fragment clusters are approximated 
   by spheres. The rotational constants $A$, $B$, and $C$ of each cluster were computed in the most stable geometry 
   described in Sect. \ref{sec:structure}. Assuming that inertia momenta should be combined linearly, the rotation 
   constant of the equivalent sphere was evaluated as the reduced value $3ABC/(AB+AC+BC)$. The monomer fragment is 
   considered as an oblate top molecule.
   
   \citet{calvo_statistical_2003} show that the need for an accurate description of the attractive interaction
   potential between the fragments increases with the rotation of the parent cluster and that the simple parametrisation 
   as $-C_6/r^6$ holds for non-rotating clusters. We found that for J=0, the KER and evaporation rates are highly 
   insensitive to the value of $C_6$ (see Appendix~\ref{anx:sensitivity}). For each cluster, therefore, we simply 
   evaluated $C_6$ by fitting the dissociation curve along the specific evaporation path for which the most external
   molecule leaves radially the cluster without rotating.

   Finally, the errors resulting from the simple description of the geometry and the long-range interaction potential
   are found to be negligible with respect to the effect of varying the angular momentum (see Appendix~\ref{anx:sensitivity}).

\subsubsection{Assessment of the statistical method}
\label{sec:PST_assessment}
   
\begin{figure}
      \begin{center}
         \includegraphics[angle=270, width=0.49\textwidth, keepaspectratio]{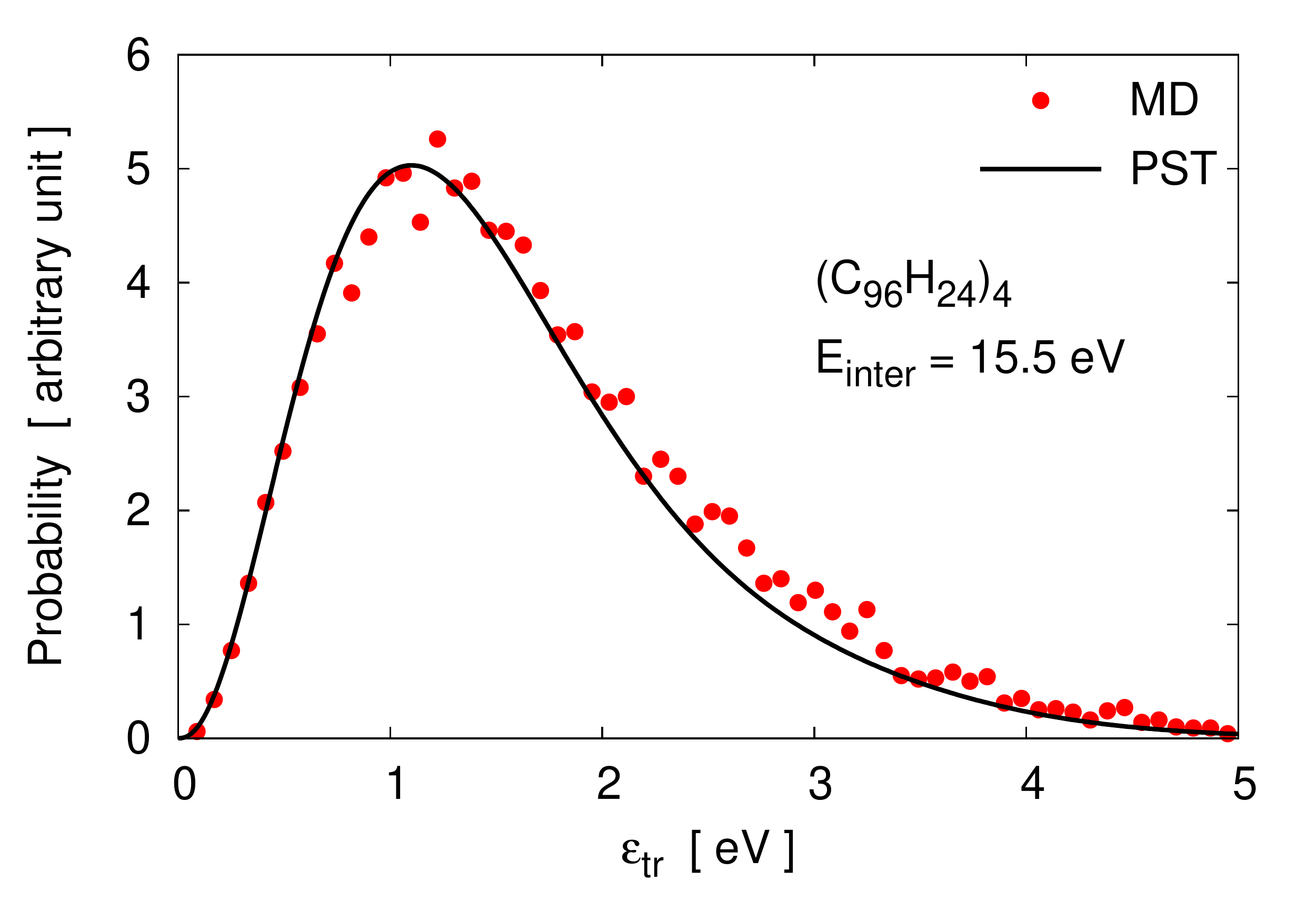}
      \end{center}
      \caption{Distribution of the kinetic energy released upon the evaporation of \agr{96}{24}{4}{} with an intermolecular 
      energy equal to 15.5 eV prior to evaporation. The points represent the results from MD and the solid line the 
      prediction based on PST.}
      \label{fig:KER}
   \end{figure}

   \citet{weerasinghe_absolute_1993} have shown the relevance of the KER distributions for assessing the accuracy of the 
   statistical calculations. For each MD evaporative trajectory, the KER was
   recorded to build a histogram that was compared to the distribution predicted by PST. Figure~\ref{fig:KER} shows a 
   typical KER for the example of \agr{96}{24}{4}{}. Despite the crude approximations concerning the geometry and the 
   long-range potential parameter of the clusters, MD and PST results exhibit a comparable level of agreement as 
   reported by other authors for similar studies \citep{rapacioli_formation_2006,calvo_accurate_2010}.
   
   An alternative assessment comes from the relative evolution of evaporation rates $\kevap^{\text{rigid}}$ as derived from MD 
   simulations and PST. We tuned the coefficient $C_0$ in Eq.~(\ref{eq:kevap-rigid}) to best fit the MD results. Most MD 
   points deviate by less than 15\% from the calibrated PST rates, and the maximum deviation is less than 60\%. Compared 
   with the orders of magnitude spanned by the evaporation rates, these moderate differences reinforce the validity of 
   this method.

\subsubsection{Absolute evaporation rates}
\label{sec:PST_results}

\begin{figure*}
      \begin{center}
         \includegraphics[angle=270, width=0.49\textwidth, keepaspectratio]{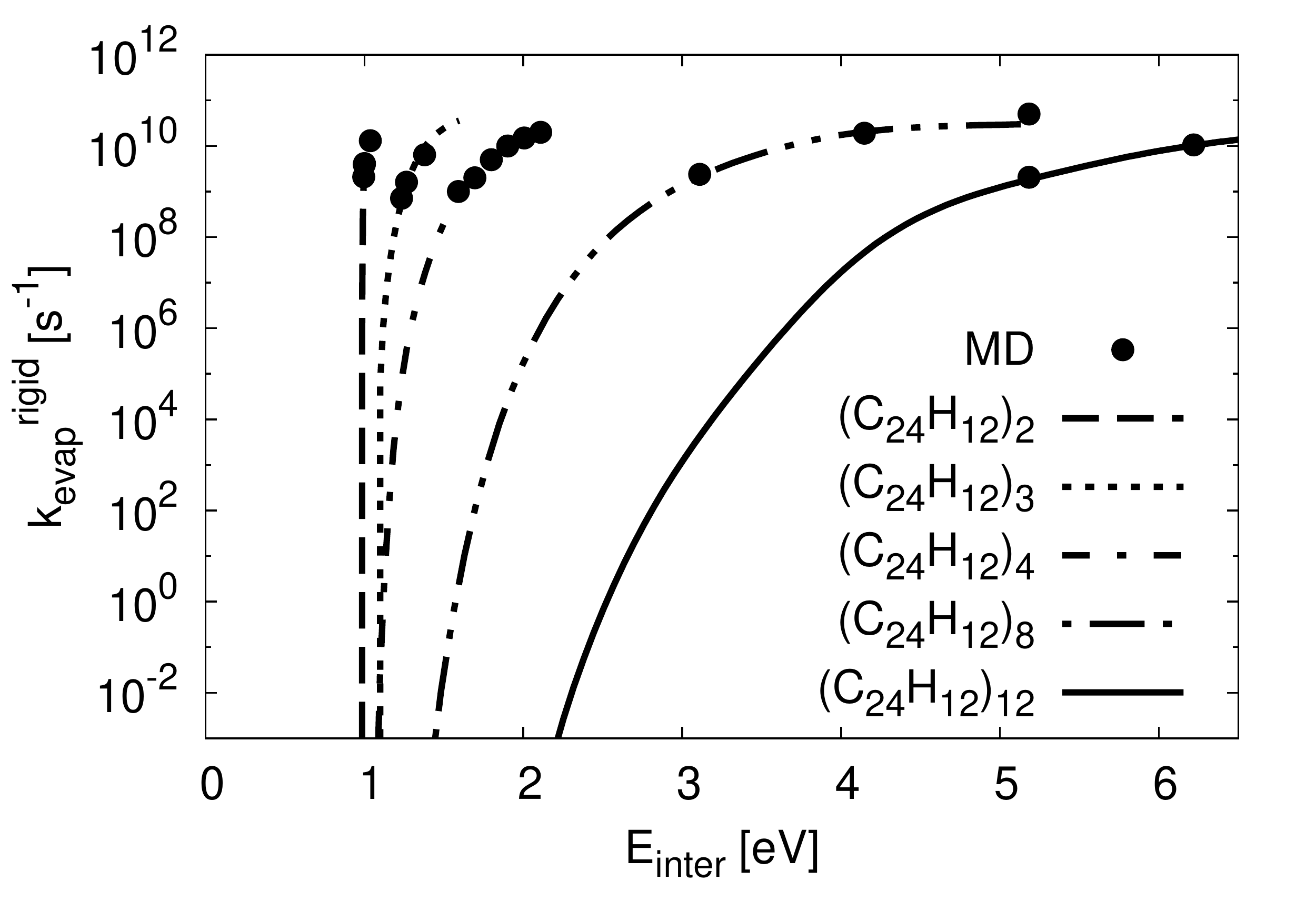}
         \includegraphics[angle=270, width=0.49\textwidth, keepaspectratio]{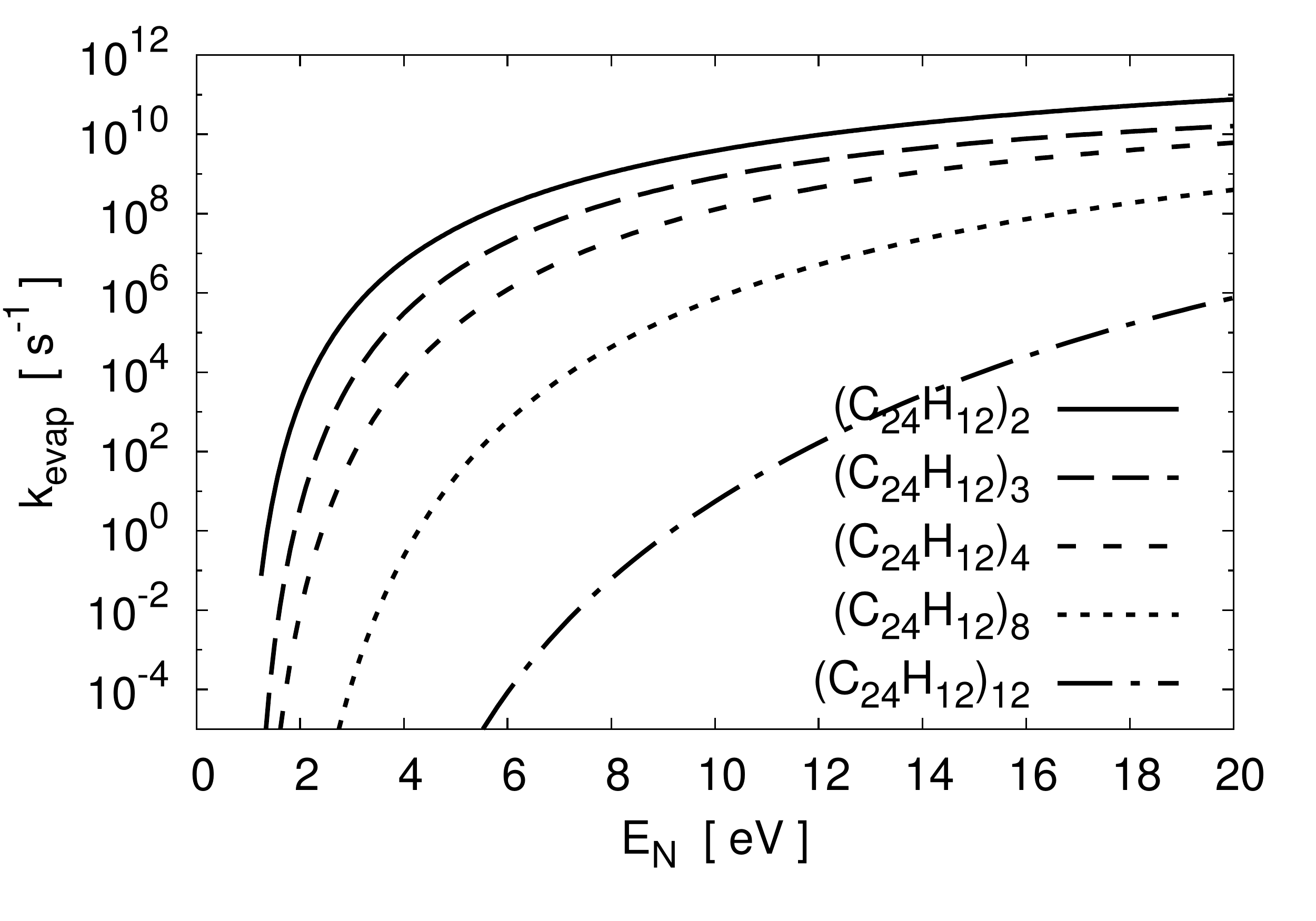}
      \end{center}
      \caption{{\it Left}: Evaporation rates of some coronene clusters as a function of the internal 
      energy in the rigid- molecule approximation from PST (lines) and MD (points) calculations. 
      {\it Right}: corresponding evaporation rates as a function of the total energy taking the 
      distribution of energy over the inter- and intra-molecular modes into account.}
      \label{fig:kevap_coro}
   \end{figure*}

   As mentioned in the previous section, the relative rates $\kevap^{\text{rigid}}$ predicted by PST were calibrated 
   with the absolute results from MD simulations. 
   The resulting absolute evaporation rates $\kevap^{\text{rigid}}$ are summarised in Fig.~\ref{fig:kevap_coro}.
   The size of the molecules affects the results essentially through the increase in the dissociation energy 
   with the number of carbon atoms and, similarly to the VDOS (Fig.~\ref{fig:scaledVDOS}), the curves roughly overlap 
   when plotted as a function of $\Einter / D_0$. When increasing the number of molecules, the intermolecular energy is
   distributed amongst more intermolecular modes, leading to less steep rates. 
   
\begin{figure}
      \begin{center}
         \includegraphics[angle=270, width=0.49\textwidth, keepaspectratio]{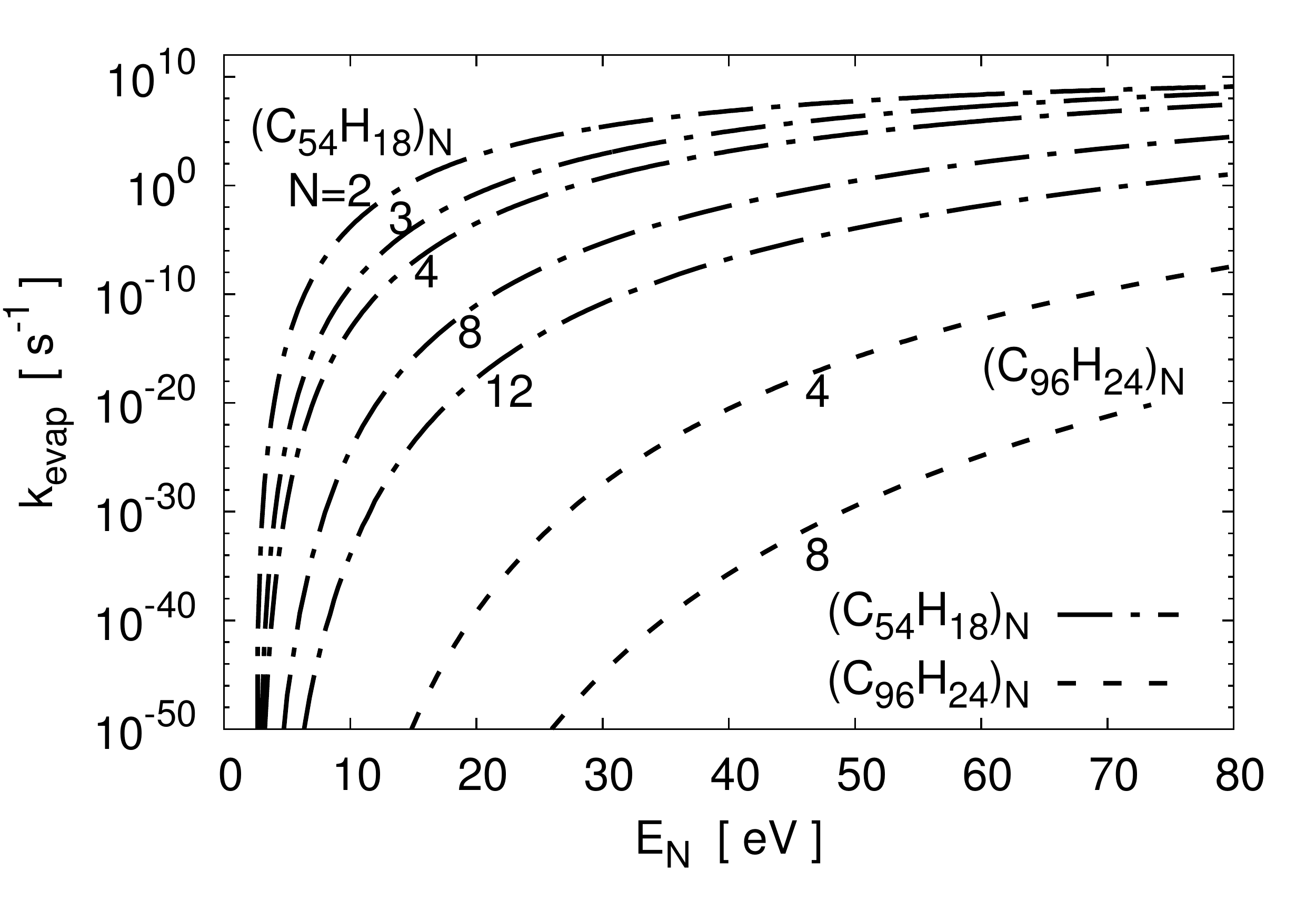}
      \end{center}
      \caption{Same as in the right panel of Fig.~\ref{fig:kevap_coro} for clusters of circumcoronene (dot-dashed lines) 
      and circumcircumcoronene (dashed lines).}
      \label{fig:kevap_Ccoro}
   \end{figure}

   The absolute values of the evaporation rates $\kevap$ as a function of the total energy were derived from 
   Eq.~(\ref{eq:kevap-link}). No further calibration is needed for this procedure. The results are presented in 
   Figs.~\ref{fig:kevap_coro} and \ref{fig:kevap_Ccoro}. The rates $\kevap$ increase much more slowly with energy than 
   $\kevap^{\text{rigid}}$ because they also take the distribution of the energy over the intramolecular modes into account. Their 
   evolution with the total number of carbon atoms in the cluster is faster when increasing the size of molecules than 
   when increasing their number. This results from the combination of two effects: the dissociation energy increases 
   with the size of the molecules, and for a given number of carbon atoms in the cluster, the fraction of intermolecular 
   modes decreases with the size of molecules. 
   
   The simpler approach proposed by \citet{rapacioli_formation_2006} provides very different rates at low 
   energies since they only consider the most probable value of the intermolecular energy. The difference vanishes at high
   energies for which high values of $\Einter$ are statistically probable. Both results are compared to available 
   experimental data in the next section.

\subsection{Comparison with experimental results}
\label{sec:schmidt_exp}

   The study of the evaporation of coronene clusters performed by \citet{schmidt_coronene_2006} provides the only 
   available data for testing the robustness of our approach. Using an aggregation source, the authors generated a 
   population of neutral coronene clusters that were thermalised in a helium thermal bath, irradiated by a 4 eV pulsed 
   laser, and detected with a time-of-flight mass spectrometer. They observed ionised coronene clusters with a wide size 
   distribution up to at least 26 units. The authors also provide experimental constraints on the evaporation of neutral 
   coronene clusters that we can analyse here using the results presented in Sect.~\ref{sec:PST_results}. 
   
   The time of 500 $\mu$s required for the neutral clusters to cross the thermalisation chamber enables thermal equilibrium 
   with the helium bath to be reached. The clusters are then photoionised by the 4 eV pulsed laser and detected in the 
   time-of-flight mass spectrometer. Varying the temperature of the thermal bath, the authors observed a full 
   evaporation above 470 $\pm$ 50 K and no evaporation below.
   
   We assume that the evaporation occurs in the thermal bath. The minimum evaporative rate a cluster needs to evaporate 
   within 500 $\mu$s is typically 2000 s$^{-1}$. According to our results, such rates occur at internal energies of 2.2, 
   2.8, 3.7, 6.5, and 13.8 eV for \agr{24}{12}{N}{} with N=2, 3, 4, 8, and 12, respectively. Using the caloric curves 
   derived from the VDOS of each cluster, these energies correspond to temperatures of 494, 446, 439, 406, and 478 K, in 
   good agreement with the experimental threshold temperature of 470 $\pm$ 50 K. 
   
   The same reasoning was applied using the evaporation rates derived by \citet{rapacioli_formation_2006}, leading to 
   rather different threshold temperatures, with $\sim$800 K for \agr{24}{12}{4}{} and $\sim$600 K for 
   \agr{24}{12}{13}{}. This shows the importance of considering the distribution of internal energy over both the inter- 
   and intra-molecular modes to derive the evaporation temperature, as implemented in our study.

\section{Astrophysical implications}
\label{sec:appli_astro}
   
   To evaluate the lifetime of neutral PAH clusters in PDRs, we used the results presented in the previous section to 
   model their evolution as a function of time in the physical conditions typical of such environments. In this section, we 
   present the astrochemical model and its results, as well as some perspectives for future developments.
   
   \subsection{Modelling the evaporation of PAH clusters in photodissociation regions}\label{sec:model_astro}
   To model the time evolution of PAH clusters in PDRs, we used the model presented in \citet{montillaud_evolution_2013} 
   that we have developed to study the charge and hydrogenation states of PAHs in interstellar conditions. This model 
   fully describes the time evolution of the abundance and the internal energy of the species, based on their evaporation 
   and IR emission rates. It thereby naturally considers the building up of internal energy due to possible 
   successive absorptions of several UV photons before the species have completely cooled down by IR emission 
   (multi-photon events). Such events are expected to be crucial considering the low values of the evaporation rates of 
   circumcoronene and circumcircumcoronene clusters at 13.6 eV, which corresponds to the maximum energy that can be 
   absorbed via a single UV photon. In the following sections, we explain the hypotheses of the model, and show how 
   we computed the complementary molecular data needed for the modelling.
   
   \subsubsection{General hypotheses of the astrochemical model}\label{sec:model_astro}
   The numerical aspects of the astrochemical model are presented in detail in \citet{montillaud_evolution_2013}. We 
   assumed that the clusters can evolve according to the following processes:
   \begin{itemize}
      \item gain of one or several molecules by collision with a PAH or another PAH cluster,
      \item increase in the internal energy by absorption of UV photons,
      \item decrease in the internal energy by emission of IR photons,
      \item loss of one molecule by evaporation.
   \end{itemize}
   We modelled the clusters of coronene, circumcoronene, and circumcircumcoronene separately. We limited the number of
   molecules in PAH clusters to 12 molecules. Initially, the abundances of the clusters with $N$ molecules were set to 1 
   for $N=12$ and to 0 otherwise. 

   \subsubsection{Formation of PAH clusters}\label{sec:formation_astro}
   \citet{rapacioli_formation_2006} have shown that for temperatures below a few 1000 K the collision of coronene 
   molecules (\ch{24}{12}) always leads to the formation of a dimer. The situation is even more favourable for 
   collisions between larger molecules, between a PAH molecule and a cluster, or between two clusters, because of the
   larger number of degrees of freedom available to distribute the collisional energy. Therefore, we assumed a sticking 
   coefficient S=1, and computed the collision rates using the average geometric cross-sections. We show in 
   Sect.~\ref{sec:results_astro} that the results do not depend on this assumption.

   \subsubsection{Interpolation of $\kevap$}\label{sec:completing_kevap}
\begin{figure*}
      \begin{center}
         \includegraphics[angle=270, width=0.49\textwidth, keepaspectratio]{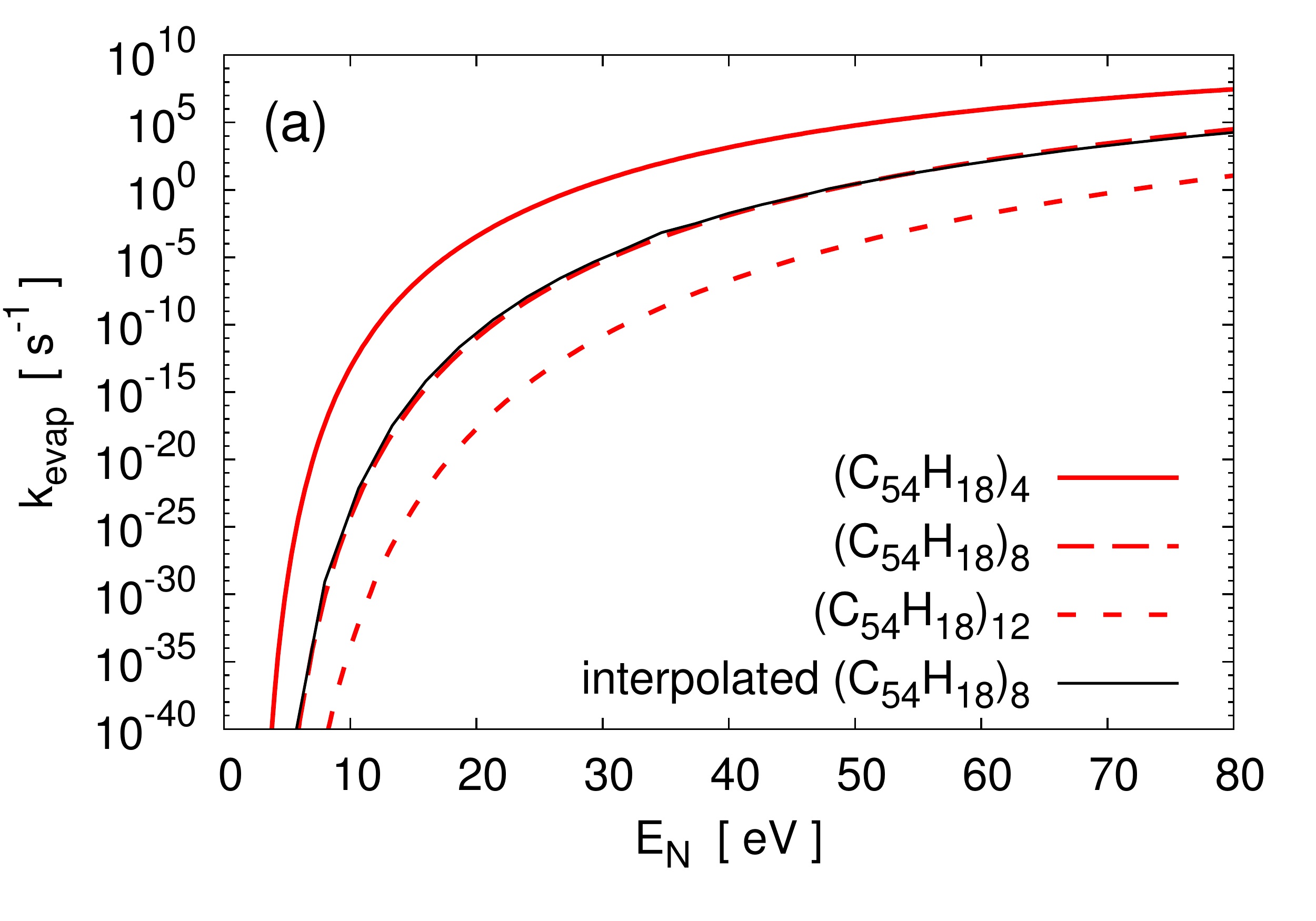}
         \includegraphics[angle=270, width=0.49\textwidth, keepaspectratio]{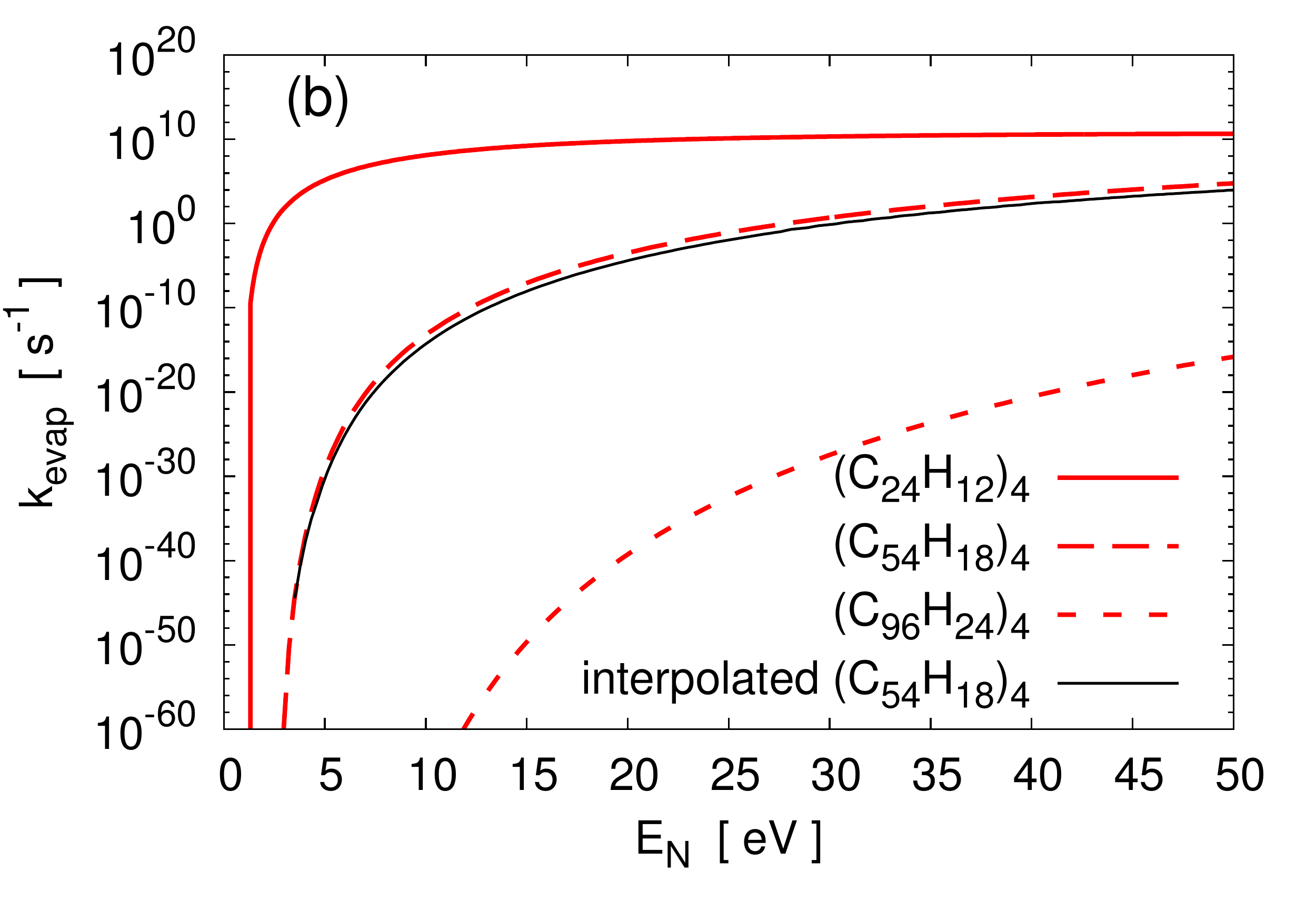}
      \end{center}
      \caption{Comparison of the curves of $\kevap$ obtained from full statistico-dynamical calculations (red curves) or
      from the interpolation methods (black curves) for clusters made of the same molecules (a) or for clusters with the
      same number of molecules (b).}
      \label{fig:interpolation}
   \end{figure*}

   In Sect.~\ref{sec:result_disc}, we computed the evaporation rates for a limited set of PAH clusters, containing
   2, 3, 4, 8, and 12 molecules for coronene and circumcoronene clusters, and only 4 and 8 molecules for 
   circumcircumcoronene clusters. The computed evaporation rates show very regular trends with the size and number of 
   molecules in the cluster. Figure~\ref{fig:interpolation}(a) shows the result of the logarithmic interpolation between 
   the curves of evaporation rates of \agr{54}{18}{4}{} and \agr{54}{18}{12}{} to evaluate the evaporation rate of
   \agr{54}{18}{8}{}. The result is in good agreement with the rate obtained by the detailed calculation presented in 
   Sect.~\ref{sec:result_disc}. Therefore, we have built a full set of evaporation rates for coronene, circumcoronene, 
   and circumcircumcoronene clusters using the equation
   \begin{equation}
      \label{eq:kevap_interpo_Npah}
      \log(\kevap^n) = (1-x) \log(\kevap^m) + x \log(\kevap^p),
   \end{equation}
   where $x=(n-m)/(p-m)$, $m$ and $p$ are the numbers of molecules in the clusters with known evaporation rates 
   $\kevap^m$ and $\kevap^p$, and $n$ is the number of molecules in the cluster for which the evaporation rate is 
   interpolated.

   Interestingly, we find similar trends with the size of molecules in the clusters. Figure~\ref{fig:interpolation}(b)
   shows that one can derive a fair estimate of the evaporation rate of \agr{54}{18}{4}{} from the interpolation of the
   evaporation rates of \agr{24}{12}{4}{} and \agr{96}{24}{4}{} using the formula
   \begin{multline}
      \label{eq:kevap_interpo_Nc}
      \log(\kevap^n(E/D_{0,n})) = \\ (1-x) \log(\kevap^m(E/D_{0,m})) + x \log(\kevap^p(E/D_{0,p})),
   \end{multline}
   where $D_{0,i}$ ($i=n$, $m$, $p$) is the dissociation energy of the species $i$. 

   These regular trends stem from the very similar structures of the clusters. Indeed, for coronene clusters, we find 
   that the interpolation provides poor estimates when one bases the interpolation on one cluster smaller than eight
   molecules and the other larger than eight molecules. This is related to the transition from the one-stack to two-stack 
   geometry of this cluster. One should therefore use this interpolation method with caution, and only for clusters with 
   similar geometries.

   \subsubsection{IR emission rates}\label{sec:kIR}
   Using our astrochemical model requires computing the cooling rate for each cluster, which is dominated by the 
   emission of IR photons from intramolecular vibrational modes. We follow \citet{rapacioli_formation_2006} by considering 
   that the vibrational properties of the molecules involved in a cluster are weakly perturbed by the other molecules of 
   the clusters. Therefore, the IR emission rate $\kIR$ of a cluster as a function of the total intramolecular energy in 
   the cluster $\Eintra$ is scaled from the $\kIR$ of the monomer
   \begin{equation}
      \kIR(\Eintra) = N \times k_{\rm IR, monomer}(\Eintra/N),
   \end{equation}
   where $N$ is the number of molecules in the cluster. The intramolecular energy is assumed to be equally distributed 
   among the molecules. The error resulting from this approximation on $\kIR(\Eintra)$ is expected to be
   small because this rate is mainly determined by the most probable values of $\Eintra$. Furthermore, the final results, 
   i.e. the time evolution of clusters in astrophysical conditions, are quite insensitive to the exact value of $\kIR$, 
   as noticed on comparable systems by \citet{le_page_hydrogenation_2001} or \citet{montillaud_evolution_2013}, among others.
   In contrast, it is necessary to properly describe the distribution of the total energy $\Etot$ between $\Eintra$ and 
   $\Einter$. We computed the IR emission rate as a function of the total internal energy $\Etot$ consistently with the 
   method presented in Sect.~\ref{sec:rigid} using
   \begin{equation}
      \kIR(\Etot) = \int_{0}^{\Etot} {\cal P}(\Eintra) \times N \times k_{\rm IR, monomer}(\Eintra/N) d\Eintra, 
   \end{equation}
   where ${\cal P}(\Eintra)$ is the complementary probability of ${\cal P}(\Einter)$. The resulting values of $\kIR$ 
   increase slowly with $\Etot$ in the 1-100 s$^{-1}$ range. 
   
   \subsection{Results}\label{sec:results_astro}
   We computed the time evolution of the three kinds of clusters for a hydrogen nuclei density of $n_{\rm H}=1.5 \times 
   10^{3}$ cm$^{-3}$ and a gas temperature of $T=100$ K. The UV radiation field was scaled from the interstellar
   radiation field of \citet{mathis_interstellar_1983} to values ranging from $G_0=1$ to $3\times10^7$ in Habing units
   \citep{habing_interstellar_1968}.
   
\begin{figure*}
      \begin{center}
         \includegraphics[width=0.49\textwidth, keepaspectratio]{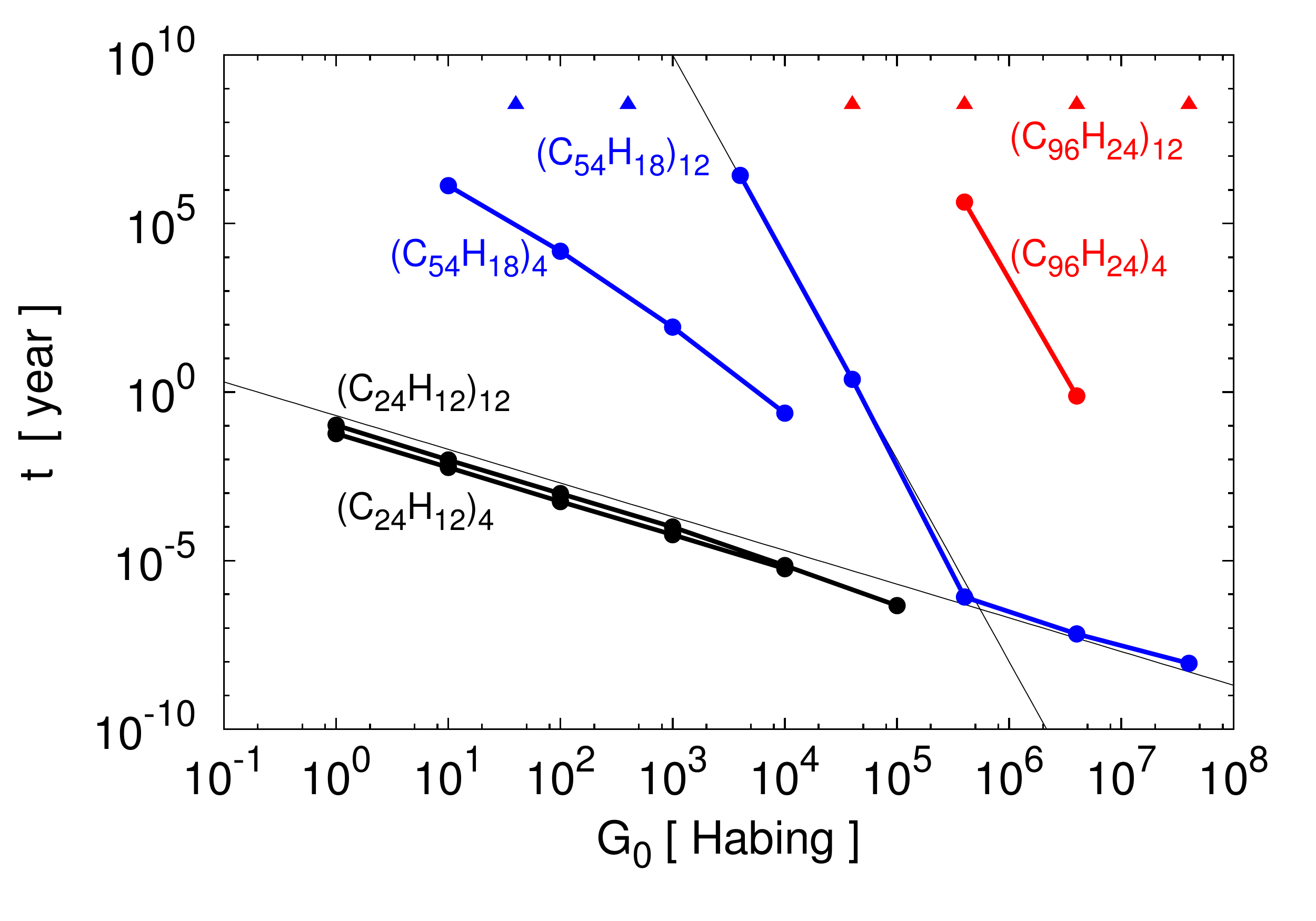}
         \includegraphics[width=0.49\textwidth, keepaspectratio]{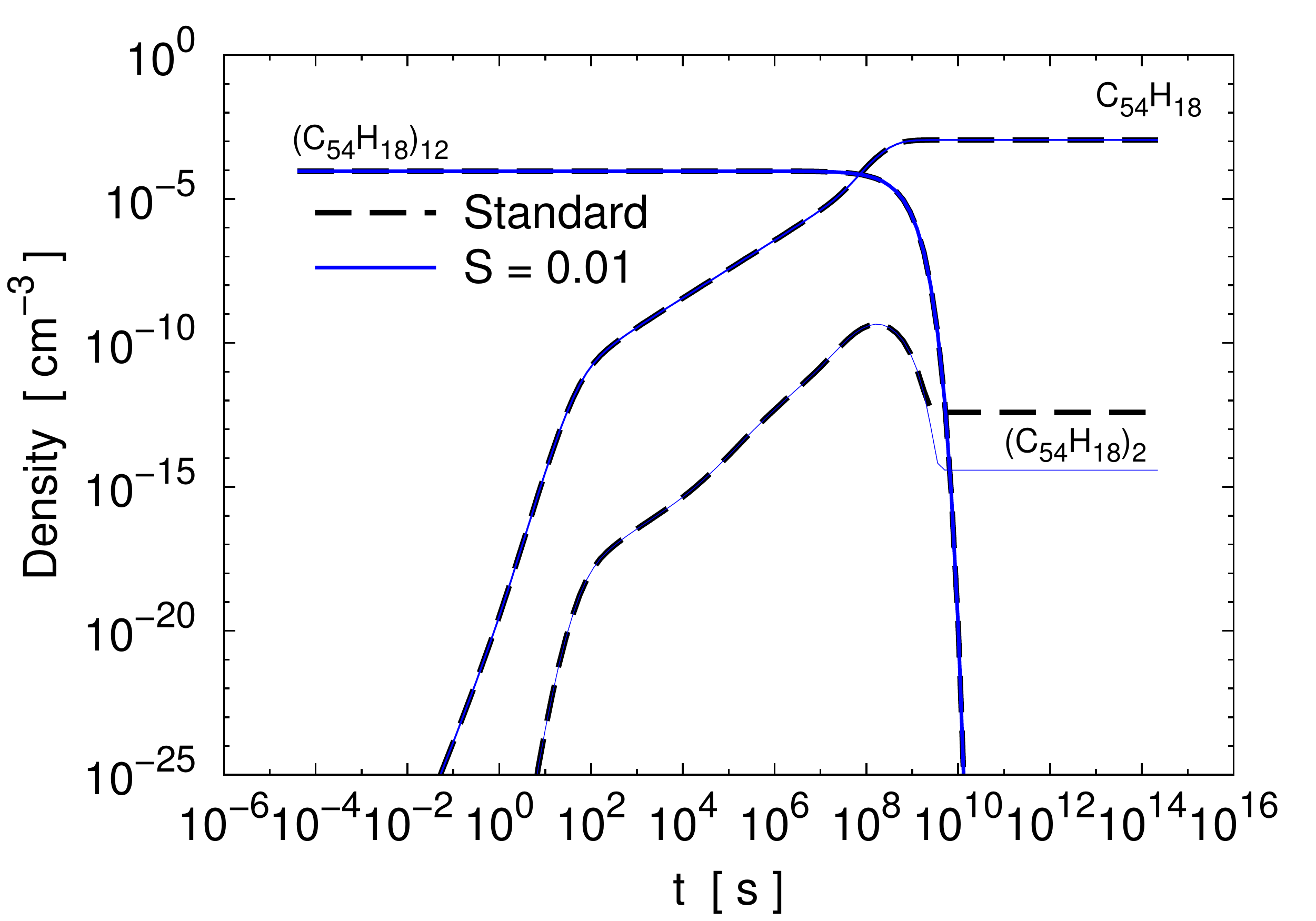}
      \end{center}
      \caption{{\it Left}:
      Evaporation timescales of a few PAH clusters as a function of the intensity of the UV radiation field, obtained 
      with a model considering the growth of clusters by collisions with a sticking coefficient S=1, a hydrogen 
      nuclei density of $n_{\rm H}=1.5\times10^3$ cm$^{-3}$, and a gas temperature of $T=100$ K. The thin black lines 
      show the slopes of -1, characteristic of single-photon evaporation, and of -6, indicative of highly multi-photon 
      evaporations.
      {\it Right}: Results of the model for a UV radiation field of $G_0=4\times10^4$ in units of Habing as a 
      function of time for circumcoronene clusters when considering the growth of clusters by collisions with S=1 (black 
      dashed lines) or S=0.01 (blue solid lines).
      }
      \label{fig:astro_results}
   \end{figure*}

   Figure~\ref{fig:astro_results} shows the variations in the timescale for photo-evaporation of a few clusters as a 
   function of the intensity of the UV-visible radiation field. Different clusters exhibit very different behaviours. 
   Coronene clusters are found to be quickly evaporated even in faint radiation fields. In contrast, 
   circumcircumcoronene clusters require very intense radiation fields to evaporate. \agr{96}{24}{12}{} does not 
   evaporate significantly on timescales of $10^8$ years, longer than the typical lifetime of a molecular cloud. In 
   between, circumcoronene clusters show an interesting range of timescales. Smaller clusters are resistant in faint 
   radiation fields but are quickly evaporated for values higher than $G_0 \approx 10^3$ in Habing units. The larger 
   circumcoronene clusters require rather intense radiation fields ($G_0 \gtrsim 10^4$ in Habing units) to evaporate. 
   Therefore, our results show that PAH clusters made of medium-sized PAH molecules ($\sim 50$ C-atoms) have evaporation 
   properties compatible with the evaporating very small grains observed in numerous photodissociation regions.
   
   The evaporation timescales of coronene clusters vary as 1/$G_0$, consistently with these clusters being evaporated by 
   a single photon, because their threshold energy $E_{\rm th}$ (defined by $\kevap(E_{\rm th}) = \kIR(E_{\rm th}$)) is 
   lower than the maximum energy carried by a single photon (13.6 eV). Conversely, evaporation timescales of 
   circumcoronene and circumcircumcoronene clusters have steeper variations with $G_0$, because multiphoton events are 
   necessary to evaporate these clusters owing to their high threshold energies. Interestingly, Fig.~\ref
   {fig:astro_results} shows in the case of \agr{54}{18}{12}{} that for very intense radiation fields, the evaporation
   timescale presents a break and varies as 1/$G_0$. This is because the mean time between the successive absorption of 
   two UV photons becomes shorter than the cooling timescale, and there is no longer any difference between a single 
   photon and a multi-photon event.
   
   We checked whether our results depend on the value of the sticking coefficient by computing one model with a sticking
   coefficient of S=0.01. Figure~\ref{fig:astro_results} shows on the example of circumcoronene clusters with 
   $G_0=4\times10^4$ in Habing units that the abundance of dimers is divided by 100 compared to the case with S=1, once
   steady state is reached. However, in these conditions, the abundance of the dimer is always dominated by at least five
   orders of magnitude by \agr{54}{18}{12}{} (on short timescales) or \ch{54}{18} (on long timescales), while the 
   abundance and the timescales of the dominant species do not change significantly. Our results are thus robust 
   against the assumptions for the growth of clusters by collisions, and this validates the approximation in 
   Sect.~\ref{sec:formation_astro}, in typical PDR conditions. 
   
   \subsection{Perspectives}\label{sec:perspectives}

   The results presented in the previous sections show that PAH clusters are promising candidates for the population of 
   astronomical eVSGs. Still, we only considered neutral homo-molecular clusters of highly symmetric compact 
   PAHs, and natural mixtures are unlikely to be so simple. We examine here a few directions for future studies.
   
   In Appendix~\ref{sec:VDOS_cl_anh}, we show that vibrational densities of states of clusters of large PAHs can be 
   derived from a simple scaling procedure from clusters of smaller molecules. This could be useful to compute easily
   evaporation rates of other clusters.
   
   The binding energies of a few heteroclusters have been evaluated by \citet{rapacioli_stacked_2005}. They find their 
   energies to be intermediate between the binding energies of clusters made of only the largest or the smallest 
   molecules. For example, \agr{24}{12}{4}{}, \agr{24}{12}{2}{}\agr{54}{18}{2}{}, and \agr{54}{18}{4}{} have binding 
   energies of ${-294.5}$, ${-572.8}$, and ${-802.0}$ kJ/mol, respectively. Similarly, less symmetric PAHs are likely to 
   deviate from the trends that are found in this paper. Therefore, considering a whole family of interstellar PAHs with 
   sizes ranging between 24 and 96 carbon atoms provides a wealth of clusters with properties that could fit the 
   observational constraints. 
   
   Other processes may also occur before clusters photo-evaporate. \citet{rapacioli_modelling_2009} show that the 
   ionization potential of coronene clusters decreases when the number of units in the cluster increases. The authors 
   report a decrease from 7.6 eV for the monomer to 6.4 eV for the octomer. In addition, larger PAHs have a lower 
   ionisation potential than smaller ones. For example, the ionization potential of \ch{96}{24} is estimated to be
   $\sim5.7$ eV \citep{ruiterkamp_pah_2005}. A description of the charge state of PAH clusters in PDRs is therefore 
   necessary. In addition, because the binding energy of ionised PAH clusters containing a small number of molecules is 
   significantly larger than in the corresponding neutral clusters \citep{rapacioli_modelling_2009}, ionisation is 
   expected to affect the photo-stability of these species in PDRs.
   
   PAHs in clusters may not be always saturated, for instance, because of interactions with cosmic rays. This could 
   imply intra-cluster reactivity and the formation of chemical bonds between the PAH units within the cluster. This 
   could represent a limitation to the applicability of our results, if such reactivity occurs on timescales shorter 
   than the evaporation timescales. Adaptation of the framework used in the present paper to intra-cluster reactivity 
   might be possible as long as the processes are statistical.

   Only the largest clusters that survive evaporation in PDRs (if they exist) could be found in HII regions where they 
   are submitted to EUV photons of $\sim$20-30 eV. In PDRs, such energies can only be attained in rare events by 
   sequential absorption of milder UV photons. We can therefore expect that the destruction of PAH clusters will be very 
   efficient in HII regions. The absorption of EUV photons is likely to open other relaxation pathways, such as 
   multiple-ionisation or intra-cluster reactivity, and therefore the description of the evolution of PAH clusters in HII 
   regions is beyond the scope of the present study.

\section{Conclusion}
\label{sec:conclusion}

   We have computed the absolute evaporation rates of non-rotating neutral PAH clusters of astrophysical interest,
   made of 2 to 12 molecules of coronene \ch{24}{12}, circumcoronene \ch{54}{18}, or circumcircumcoronene \ch{96}{24}. We
   thereby provided essential data for testing the relevance of PAH clusters as models for the so-called eVSGs whose sizes 
   span the range between $\sim 100$ to $\sim 1000$ C-atoms \citep{rapacioli_spectroscopy_2005, pilleri_evaporating_2012}.
   
   We performed the calculations using the statistico-dynamical approach proposed by \citet{weerasinghe_absolute_1993}, 
   and further analysed and developed by \citet{calvo_statistical_2003}. We proposed an extension of the method that 
   enables taking the distribution of the excess energy into account amongst all the inter- and intra-molecular modes in 
   a microcanonical framework, despite the use of the rigid molecule approximation. A successful comparison with the 
   experimental results of \citet{schmidt_coronene_2006} on coronene clusters, without any fitting parameter, gives us 
   confidence in the method.
   
   The evaporation rates exhibit very regular trends that we found to be related to the structural similarities between
   the studied species. When considering clusters of molecules with different geometries, as well as heteroclusters, one
   should expect deviations from these trends that could be interesting to consider for studying natural mixtures, such as 
   in astrophysical environments or in flames. The absolute values of the evaporation rates decrease strongly with 
   the number of molecules in the cluster, and even more with the size of molecules, indicating that clusters of large 
   PAHs are more likely to survive in PDRs. These results also indicate that large PAHs could play a significant role in 
   the nucleation process of soot particles.

   We evaluated the impact of most approximations that have been made, on our results. The sensitivity of evaporation 
   rates to the accuracy of the geometric and the long-range interaction parameters used in PST should be 
   considered of second order compared to the sensitivity to other approximations made in the calculations. Rather small 
   errors result from the use of our rates to PAH clusters with rotational temperatures up to a few hundred K. For 
   higher temperatures, the initial angular momentum of parent clusters should be considered, and a more accurate 
   evaluation of the geometric and long-range interaction parameters could then be necessary \citep
   {calvo_statistical_2004}.
   
   The anharmonicity of intramolecular vibrations was estimated to play a minor role thanks to the dispersion of energy
   among numerous intra-molecular modes. On the contrary, the anharmonicity of intermolecular modes is crucial for 
   reproducing the variations in the evaporation rates with excess energy revealed by molecular dynamics. Only the 
   anharmonicity resulting from the coupling between inter- and intra-molecular modes cannot be modelled with our method. 
   A comparative study with full-atom methods could help quantify the effects of this coupling on the statistical and 
   dynamical properties of molecular cluster evaporation.

   We used our molecular data to study the evolution of PAH clusters in conditions typical of PDRs. We showed that the 
   whole range of photostability properties, from very easily photo-evaporated to extremely photostable, is covered when 
   considering clusters made of PAHs with sizes in the expected range for astrophysical PAHs. Our results therefore 
   reinforce the idea that PAH clusters are good candidates for eVSGs, if the size of individual units is large enough 
   ($\gtrsim 50$ C-atoms).
   
   This work provides most of the elements needed to investigate the survival of neutral PAH clusters in the interstellar 
   medium. Nevertheless, numerous studies are needed to have a complete set of data that would enable modelling 
   realistic interstellar PAH clusters, including the effects of the charge state, the blending of molecules, and the 
   rotation of the clusters. Apart from the molecular data, valuable progress in our understanding of the nature of 
   eVSGs now relies on a closer comparison of the results of modelling and astronomical observables.
   
   \begin{acknowledgement} 
      The authors gratefully acknowledge A. Simon, who kindly provided the results of DFT calculations presented in 
      this work, and thank F. Spiegelman, M. Rapacioli, and P. Parneix for stimulating discussions and helpful comments. \

      J. Montillaud acknowledges the support of the French Agence Nationale de la Recherche (ANR), under grant GASPARIM 
      "Gas-phase PAH research for the interstellar medium", ANR-2010-BLANC-0501.
   \end{acknowledgement}

\bibliographystyle{aa}
\bibliography{Clusters}

\Online 

\appendix

\section{Calculations of vibrational densities of states}
\subsection{Classical anharmonic VDOS of intermolecular modes}
   \label{sec:VDOS_cl_anh}

   When a cluster contains enough energy to evaporate, the system abundantly explores regions of the intermolecular 
   potential surface that strongly deviate from the quadratic approximation and that is likely to experience isomerisation, 
   especially for large numbers of molecules and for small molecules. Anharmonicity of the intermolecular modes is 
   thus expected to play a significant role in the evaporation process.
   
   \citet{labastie_statistical_1990} developed the multiple histogram method to compute VDOS on the basis of MD 
   simulations. We adapted this method to our microcanonical systems following \citet{calvo_configurational_1995}. For
   this purpose, we performed additional MD simulations using the code presented in Sect.~\ref{sec:MD}. For each cluster, 
   trajectories were integrated over $10^{-9}$s for $\sim$ 30 different initial energies up to the dissociation energy. 
   When an evaporation event was detected, the last records were deleted until the velocity vector of the evaporating 
   molecule was found to point towards the fragment cluster, and a new trajectory was started with new random initial 
   velocities. The first 10000 points were systematically discarded to ensure thermal equilibrium. The recorded 
   potential energies were distributed among the 1000 bins of the histograms that we treated with a least square 
   minimisation inspired from \citet{weerasinghe_absolute_1993}.
   
\begin{figure}
      \begin{center}
         \includegraphics[angle=270, width=0.49\textwidth, keepaspectratio]{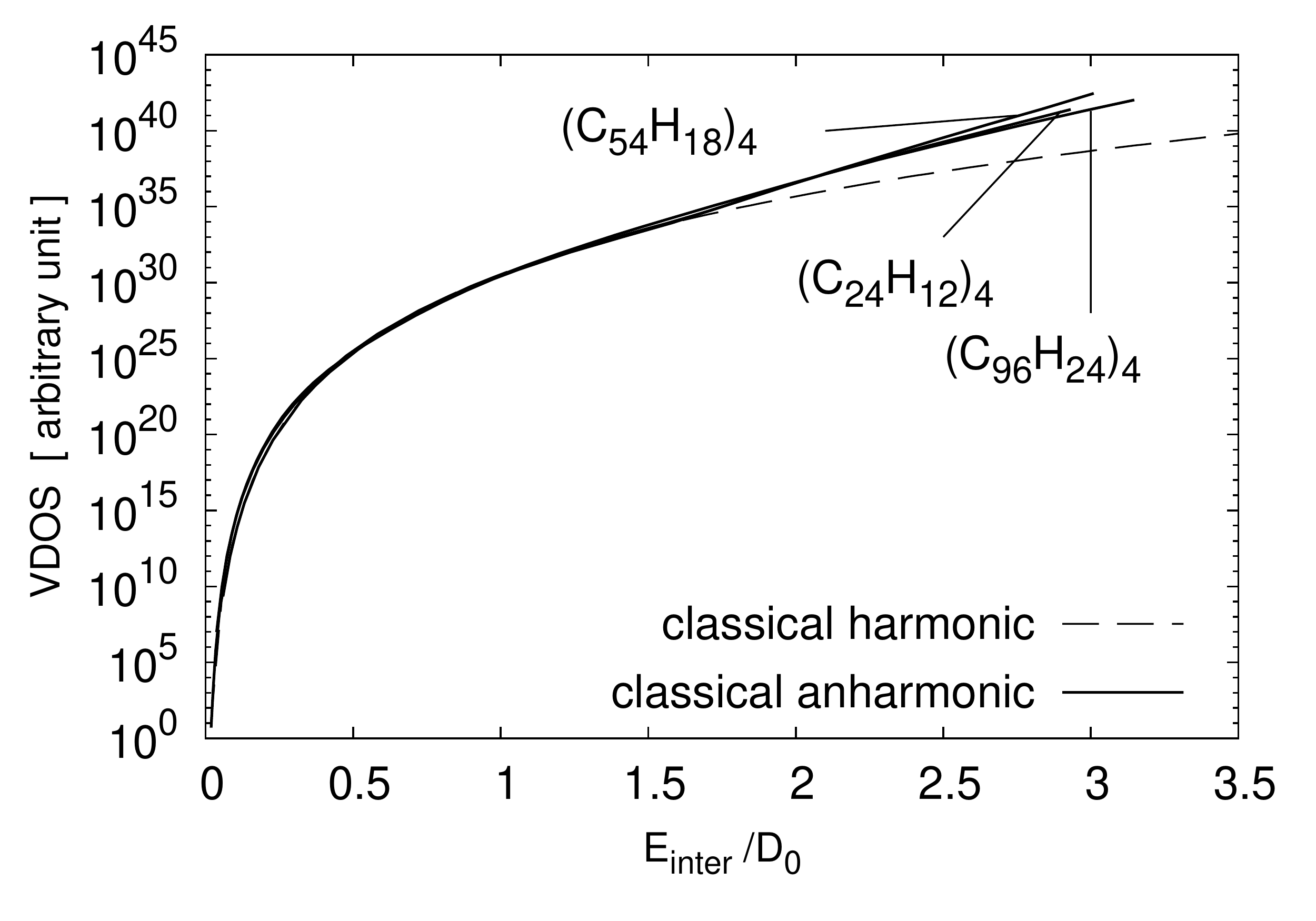}
      \end{center}
      \caption{Classical vibrational densities of states extracted from molecular dynamics simulations of the tetramers 
      of coronene, circumcoronene, and circumcircumcoronene scaled to their dissociation energy $D_0$. They exhibit 
      very similar behaviour even for energies higher than $1.7 D_0$, for which the anharmonic contribution cannot be 
      neglected.}
      \label{fig:scaledVDOS}
   \end{figure}

   The anharmonicity is found to occur at rather high energies, typically greater than 50\% of the binding energy, 
   except for the two-stack clusters \agr{24}{12}{11}{} and \agr{24}{12}{12}{}. These clusters exhibit strong 
   anharmonicity as soon as $\Einter$=2 eV, suggesting that they are more prone to isomerisation than one-stack 
   clusters. Interestingly, the intermolecular VDOS of different one-stack clusters with the same number of molecules 
   almost overlap when scaled to the corresponding dissociation energy, as illustrated with the example of tetramers in 
   Fig.~\ref{fig:scaledVDOS}. This opens the possibility of building empirical laws parametrised with the number and size 
   of PAH molecules in the cluster.

   In the following, we use this method to compute $\Ominter_{N}$ and $\Ominter_{N-1}$ in Eq.~(\ref{eq:kevap-rigid}).

\subsection{Quantum vibrational densities of states}

   The exact quantum harmonic vibrational densities of states are easily computed from the vibrational frequencies of 
   the species using the Beyer-Swinehart algorithm as proposed in \citet{stein_accurate_1973}. Density functional theory 
   (DFT) was used to compute vibrational frequencies. The functional M06-2X \citep{zhao_density_2008} was selected, 
   because it is more appropriate to describe medium-range correlation energy, such as van der Waals attraction, than 
   the popular B3LYP. It was associated to the basis 6-31G* to compute the quantum harmonic frequencies of \ch{24}{12}, 
   \agr{24}{12}{2}{}, \agr{24}{12}{4}{}, \ch{54}{18}, and \agr{54}{18}{2}{} (A. Simon, private communication).
   We applied correction factors of 0.948 to the C-H stretch mode frequencies and 0.974 for all other modes, which were 
   determined using the experimental data of \citet{joblin_infrared_1994} for coronene in neon matrix at 4 K. The DFT 
   frequencies of circumcircumcoronene \ch{96}{24} were taken from \citet{bauschlicher_infrared_2008}.
   
   Molecular dimers have six intermolecular modes. Only the three lowest frequency modes of both dimers were found 
   to be purely intermolecular. All other modes showed some intramolecular character, so we qualitatively looked 
   for the three modes with the largest intermolecular character. Following the analysis by \citet
   {rapacioli_vibrations_2007}, we considered modes to have a large intermolecular character when they involve large 
   scale motions in the molecule, i.e. collective motions of many atoms. A good typical example is the antisymmetric bowl 
   mode of the coronene dimer, as illustrated in the Fig.~1 of \citet{rapacioli_vibrations_2007}. In contrast, the {$\rm C-H$} 
   stretching mode in PAH molecules can be considered as an almost pure intramolecular mode. The three modes with large 
   intermolecular components that we selected are the antisymmetric bowl mode and two antisymmetric butterfly-like modes.
   They fall in the typical frequency range of soft modes ($\sim 50 - 120$ cm$^{-1}$).
   Together with the three purely intermolecular modes, they constitute the set of six (approximately) intermolecular 
   modes used for calculating intermolecular VDOS. Changing the chosen soft modes was found to affect the 
   corresponding VDOS only by an overall factor for energies over a few 0.1 eV. 
   
   For a cluster containing $N$ coronene or circumcoronene molecules, a set of intermolecular modes was built by 
   duplicating $N-1$ times the intermolecular modes of the corresponding dimer, and a set of intramolecular modes by 
   duplicating $N$ times the intramolecular modes of the monomer. For circumcircumcoronene clusters, we used the same 
   intermolecular modes as for circumcoronene. 

\begin{figure}
      \begin{center}
         \includegraphics[angle=270, width=0.49\textwidth, keepaspectratio]{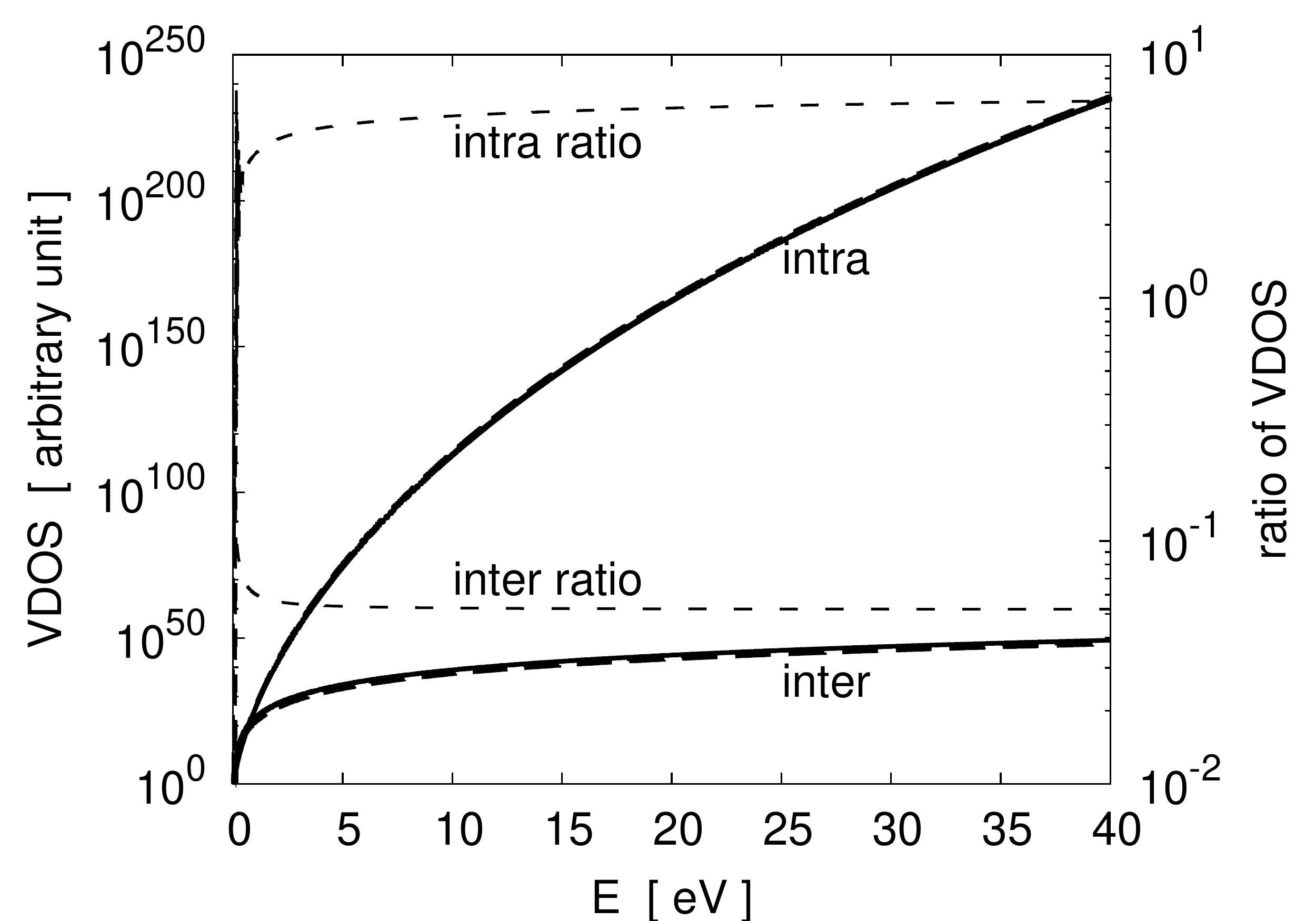}
      \end{center}
      \caption{Inter- and intra-molecular vibrational densities of states of \agr{24}{12}{4}{} computed using either the
      full set of DFT frequencies of \agr{24}{12}{4}{} (solid lines) or duplicated modes (long dashed lines, see text 
      for details). Their ratio is plotted on the right axis (short dashed lines).}
      \label{fig:duplic}
   \end{figure}

   We investigated the impact of this method by comparing the VDOS calculated using the DFT modes of the tetramer of 
   coronene and the VDOS computed with duplicated modes as described above. As shown in Fig.~\ref{fig:duplic}, the
   ratio of the two VDOS is constant for energies greater than $\sim$2 eV and remains within a factor of 2 below 2 eV. 
   These differences can be considered as small with respect to the numerous orders of magnitudes spanned by the VDOS. 
   In addition, the main use of these results consists in a convolution product of inter- and intra-molecular VDOS (see 
   Sect.~\ref{sec:distribs}), which is notably insensitive to these errors.
      
   A first estimation of the factor to apply to the quantum harmonic VDOS to correct for anharmonic effects was 
   empirically derived from the results of \citet{basire_quantum_2008}. According to their Fig.~2, the ratio of the full
   coupled anharmonic VDOS to the harmonic VDOS of the monomer naphthalene molecule \ch{10}{8} is approximated well by 
   $\exp (E/a)$ with $a=6$ eV. We take the dilution of the energy into account amongst the N vibrational modes of the 
   studied species by $\exp[(48/N) \times E /a]$, where 48 is the number of vibrational modes of naphthalene.
   This leads to a decreasing correction factor with increasing cluster size for a given energy. 
   In the absence of a better estimate, we apply this correction in the next section for both the inter- and 
   intra-molecular modes in order to get an estimate of the impact of anharmonicity on calculating the 
   distribution of intermolecular energy. One should note, however, that \citet{basire_quantum_2008} did not consider
   intermolecular vibrations in their calculations.

   \begin{figure}
      \begin{center}
         \includegraphics[angle=270, width=0.49\textwidth, keepaspectratio]{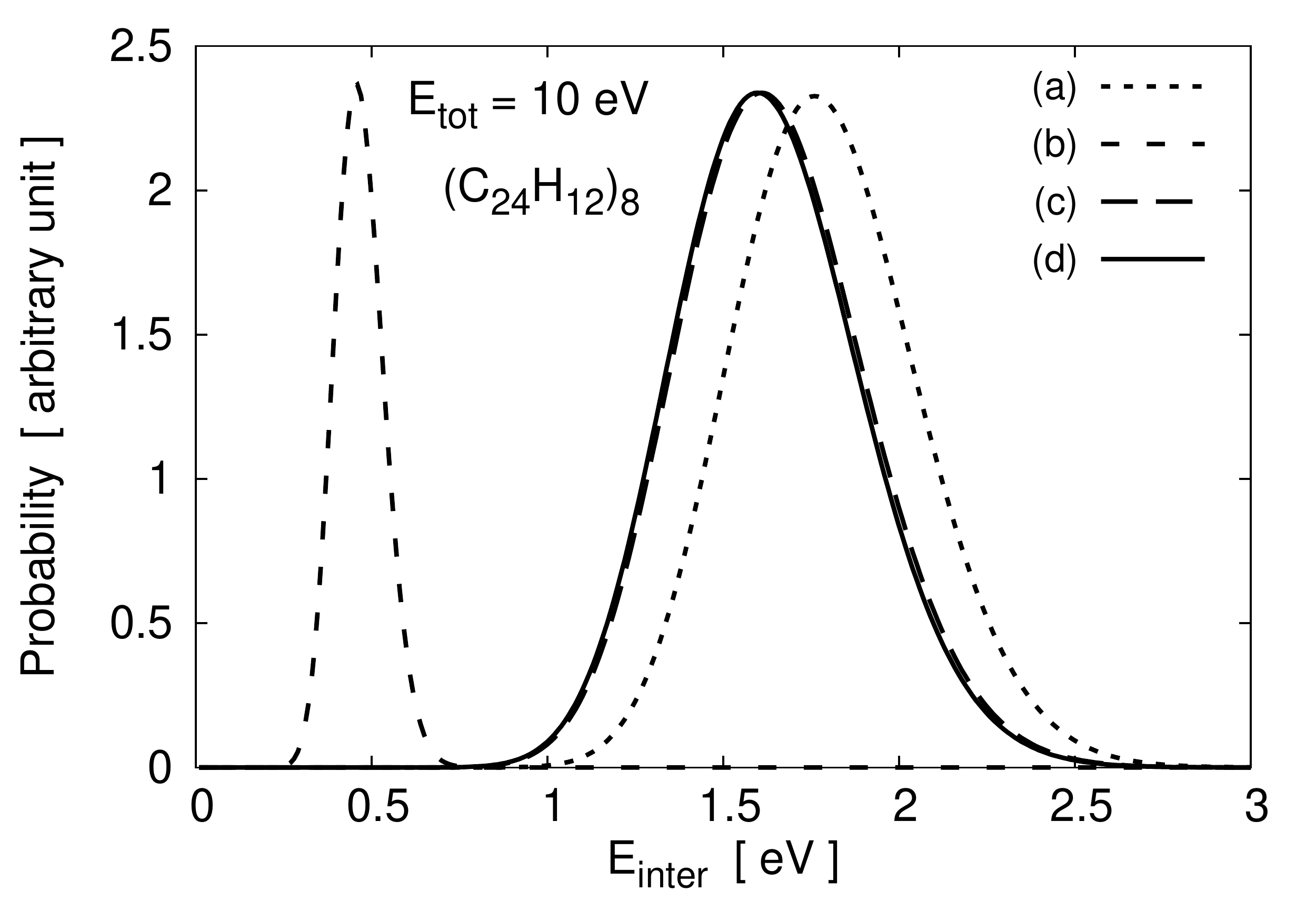}
      \end{center}
      \caption{Probability distribution of the intermolecular energy of \agr{24}{12}{8}{}, computed for several
      levels of description of the vibrational densities of states. For (a), we used different approximations for 
      intermolecular modes (classical harmonic VDOS) and intramolecular modes (quantum harmonic VDOS), but not for the 
      others: (b) classical harmonic; (c) approximated quantum anharmonic; (d) quantum harmonic.}
      \label{fig:distrib}
   \end{figure}
   
\subsection{Determining the statistical distribution of energy} \label{sec:distribs}

   The statistical distribution between inter- and intra-molecular energy is characterised by the probability 
   distribution of the intermolecular energy, computed following Eq.~(\ref{eq:distrib-Einter}). Figure~\ref{fig:distrib} 
   uses the example of \agr{24}{12}{8}{} to show that ${\cal P}_N^{\rm inter}(\Einter_N)$ peaks at rather low energies, 
   where the anharmonicity of intermolecular modes is expected to play a minor role on the peak position of 
   ${\cal P}_N^{\rm inter}(\Einter_N)$. We tested this idea by approximating quantum anharmonic VDOS using the
   anharmonic corrections derived from the work by \citet{basire_quantum_2008} as discussed in previous section. 
   Figure~\ref{fig:distrib} shows that ${\cal P}_N^{\rm inter}(\Einter_N)$ is marginally modified by the anharmonic 
   corrections. In contrast, the classical harmonic approximation leads to a significant shift in the peak position.
   In addition, since the distribution ${\cal P}_N^{\rm inter}(\Einter_N)$ is the expression of the balance between inter- 
   and intra-molecular modes, it is convenient to keep the same level of description for both kinds of modes. In 
   this study we therefore use the intermolecular energy distribution derived from the quantum harmonic VDOS for both inter- 
   and intra-molecular vibrations.

\subsection{Sensitivity of results to PST input parameters and to assumptions}\label{anx:sensitivity}

\begin{figure*}
      \begin{center}
         \includegraphics[angle=270, width=0.49\textwidth, keepaspectratio]{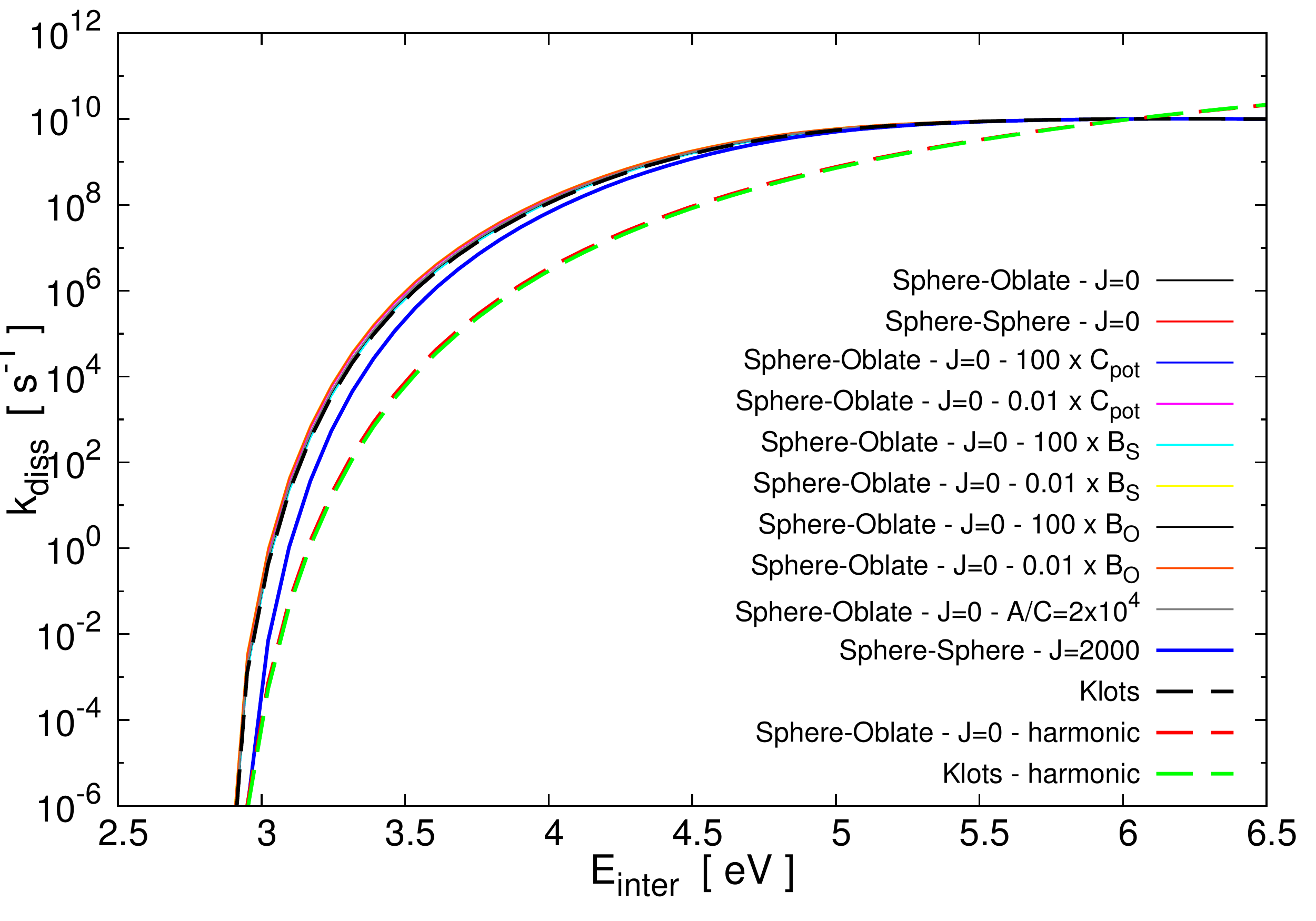}\\
         \includegraphics[angle=270, width=0.49\textwidth, keepaspectratio]{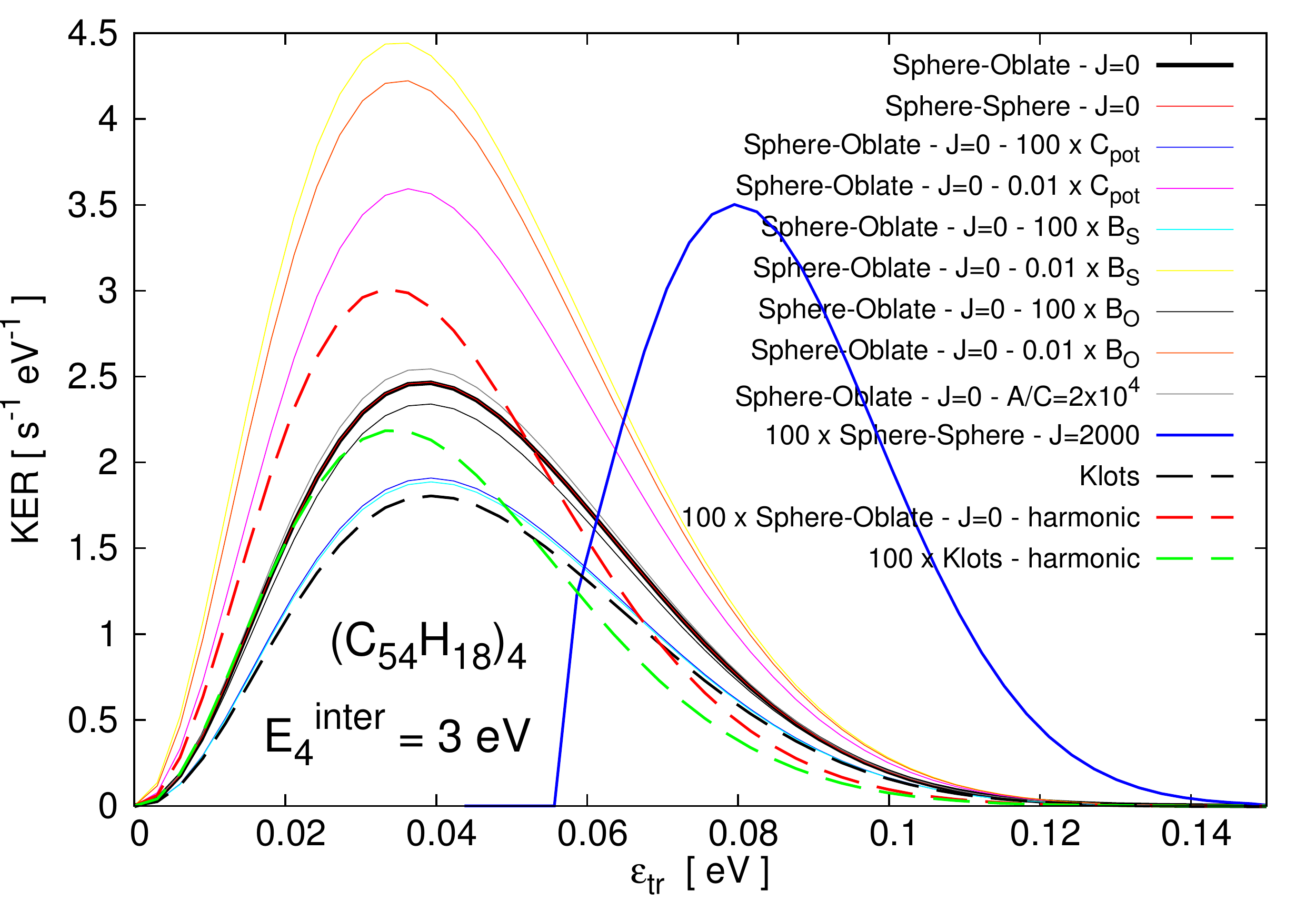}
         \includegraphics[angle=270, width=0.49\textwidth, keepaspectratio]{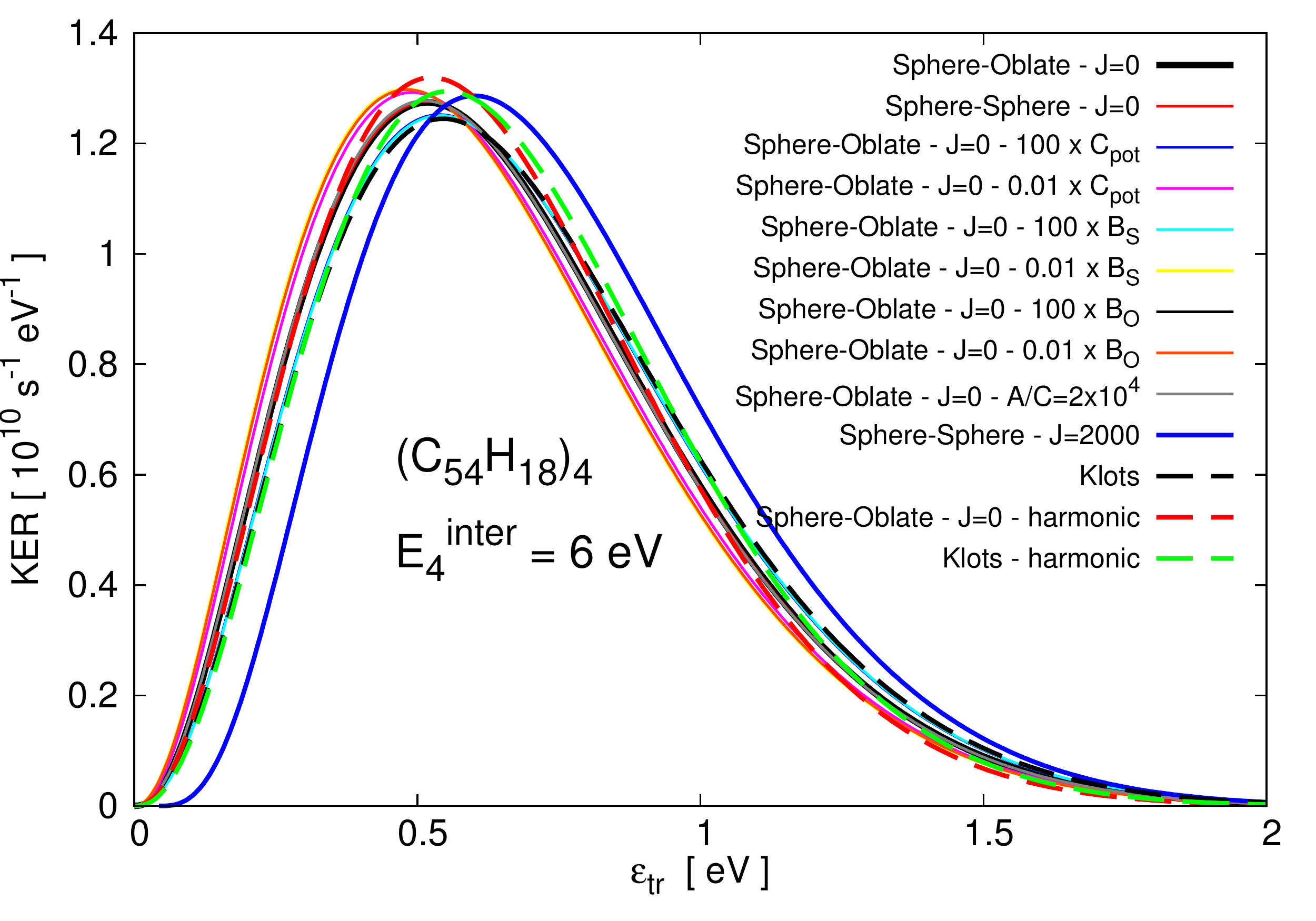}
      \end{center}
      \caption{Evaporation rate ({\it upper panel}), and distribution of KER at $\Einter_4=3$ eV ({\it lower left panel}) 
      and $\Einter_4=6$ eV ({\it lower right panel}) of \agr{54}{18}{4}{} for various combinations of parameters and 
      assumptions. $C_{\rm pot}$ is the parameter of the attractive term in the LJ potential. $B_{\rm S}$ and $B_{\rm O}$ 
      are the $B$ rotational constants of the sphere and oblate top, respectively. $A$ and $C$ refer to the first and 
      last rotational constants of the oblate top, respectively. All results are normalised to $\kevap=10^{10}$ s$^{-1}$ 
      at $\Einter_4=6$ eV. This implies that all models lead to equal areas of KER distributions at 6 eV, but different 
      ones at 3 eV. See text for details.}
      \label{fig:PST-checks}
   \end{figure*}

   We performed a series of tests to assess how sensitive our evaporation rates are to the various assumptions we made 
   and to the accuracy of the parameters. We considered the case of circumcoronene tetramer \agr{54}{18}{4}{} and used
   PST to compute the KER distributions and evaporation rate for different models. Figure~\ref{fig:PST-checks} shows the
   rigid molecule evaporation rates $k_{\rm evap}^{\rm rigid}(\Einter_N)$ and the KER distributions at $\Einter_N=3$ and 
   6 eV, defined by $ C_0 \frac{\Ominter_{N-1}(E-D_0^J-\Etr) ~~ \Gamma(\Etr,J)}{\Ominter_N(E-E_r)}$. We emphasise that
   the value of $k_{\rm evap}^{\rm rigid}(\Einter_N)$ is therefore given by the integration of the KER over $\Etr$ for
   a given $\Einter_N$ (see Eq.~\ref{eq:kevap-rigid}). For an easier comparison, we normalised all the results (i.e. we 
   chose the value of $C_0$) to have $\kevap=10^{10}$ s$^{-1}$ at $\Einter=6$ eV.
   
   We investigated the influence of the geometry by doing the computation in the sphere-sphere approximation and in the 
   sphere + oblate top approximation. In the latter case, we also did the computation for modified values of the 
   rotational constants of the sphere or the oblate top with modification factors of 100 and 0.01. In one case we 
   modified the ratio A/C by a factor $2\times10^4$. Figure~\ref{fig:PST-checks} shows that these major modifications of
   the geometry affect the KER and evaporation rates by factors of $\lesssim 3$, while the values of the evaporation 
   rates spans more than 14 decades for a total cluster energy between 3 eV and 6 eV. This very weak influence is 
   permitted by the assumption $J=0$, and higher sensitivity of the results to cluster geometry is be expected 
   when considering rotating clusters \citep{calvo_statistical_2004}.
Similar conclusions were reached when investigating the effect of the $C_6$ parameter of the attractive term in the
   Lennard-Jones potential. 
   
   We estimated the impact of the non-rotation assumption ($J=0$) by computing the KER and evaporation rate with $J=2000$.
   We found a factor of a few tens in $\kevap$ with respect to the $J=0$ case, which again should be compared to the 14 
   decades spanned by $\kevap$ between 3 eV and 6 eV of total energy. This appears to be the most critical assumption in
   our calculations.
   
   Finally Fig.~\ref{fig:PST-checks} shows the results obtained when using harmonic VDOS instead of the anharmonic VDOS 
   of Appendix~\ref{sec:VDOS_cl_anh}. The overall shape of $\kevap$ is greatly modified, illustrating that anharmonicity 
   of intermolecular modes is crucial in this study.

\end{document}